\documentclass[
   aip,
   jap,
   twocolumn,
   reprint,
   superscriptaddress,
   floatfix,
   citeautoscript,
   longbibliography,
]{revtex4-2}

\usepackage[T1]{fontenc}
\usepackage[utf8]{inputenc}

\usepackage{newtxtext}
\usepackage{newtxmath}

\usepackage{booktabs}
\usepackage{chemformula}
\usepackage{siunitx}[=v2]
\DeclareSIUnit\atom{atom}

\usepackage{graphicx}
\graphicspath{{./figures}}

\usepackage[
   unicode,
   colorlinks = true,
   allcolors = blue,
]{hyperref}

\usepackage{cleveref}

\usepackage{microtype}

\usepackage{glossaries}
\glsdisablehyper

\newacronym{1d}{1D}{one-dimensional}
\newacronym{2d}{2D}{two-dimensional}
\newacronym{3d}{3D}{three-dimensional}

\newacronym{ac}{AC}{alternating current}
\newacronym{afm}{AFM}{atomic force microscopy}
\newacronym{alc}{ALC}{avoided level crossing}
\newacronym{alcr}{ALCR}{avoided level crossing resonance}
\newacronym{api}{API}{application programming interface}
\newacronym{ariel}{ARIEL}{Advanced Rare Isotope Laboratory}
\newacronym{arpes}{ARPES}{angle-resolved photoemission spectroscopy}
\newacronym{atp}{ATP}{adenosine triphosphate}

\newacronym[sort={b-NMR}]{bnmr}{\ensuremath{\beta}-NMR}{\ensuremath{\beta}-detected nuclear magnetic resonance}
\newacronym[sort={b-NQR}]{bnqr}{\ensuremath{\beta}-NQR}{\ensuremath{\beta}-detected nuclear quadrupole resonance}
\newacronym{bca}{BCA}{binary collision approximation}
\newacronym{bcc}{BCC}{body-centred cubic}
\newacronym{bcp}{BCP}{buffered chemical polishing}
\newacronym{bcs}{BCS}{Bardeen-Cooper-Schrieffer}
\newacronym{bpp}{BPP}{Bloembergen-Purcell-Pound}
\newacronym{bsc}{BSC}{\ch{Bi2Se3:Ca}}
\newacronym{btm}{BTM}{\ch{Bi2Te3:Mn}}
\newacronym{bts}{BTS}{\ch{Bi2Te2Se}}

\newacronym{camp}{CAMP}{control and monitor program}
\newacronym{ccd}{CCD}{charge-coupled device}
\newacronym{cdw}{CDW}{charge density wave}
\newacronym{cgs}{CGS}{centimetre-gram-second system of units}
\newacronym{cmms}{CMMS}{Centre for Molecular and Materials Science}
\newacronym{codata}{CODATA}{Committee on Data for Science and Technology}
\newacronym{cpu}{CPU}{central processing unit}
\newacronym{create}{CREATE}{Collaborative Research and Training Experience Program}
\newacronym{cw}{CW}{continuous wave}

\newacronym{daq}{DAQ}{data acquisition}
\newacronym{dc}{DC}{direct current}
\newacronym{dft}{DFT}{density functional theory}
\newacronym{dos}{DOS}{density of states}
\newacronym{dqt}{DQT}{double-quantum transition}

\newacronym{efg}{EFG}{electric field gradient}
\newacronym{emim-ac}{EMIM-Ac}{1-ethyl-3-methylimidazolium acetate}
\newacronym{emim-dca}{EMIM-DCA}{1-ethyl-3-methylimidazolium dicyanamide}
\newacronym{epr}{EPR}{electron paramagnetic resonance}
\newacronym{esr}{EPR}{electron spin resonance}
\newacronym{endor}{ENDOR}{electron nuclear double resonance}
\newacronym{epics}{EPICS}{Experimental Physics and Industrial Control System}

\newacronym{fcc}{FCC}{face-centred cubic}
\newacronym{fft}{FFT}{fast Fourier transform}
\newacronym{fom}{FoM}{figure of merit}
\newacronym{fwhm}{FWHM}{full width at half maximum}

\newacronym{gga}{GGA}{generalized gradient approximation}

\newacronym{hb}{HB}{hole-burning}
\newacronym{hfqs}{HFQS}{high-field \ensuremath{Q} slope}
\newacronym{hv}{HV}{high-voltage}
\newacronym{hwhm}{HWHM}{half width at half maximum}

\newacronym{il}{IL}{ionic liquid}
\newacronym{is}{IS}{impedance spectroscopy}
\newacronym{isac}{ISAC}{isotope separator and accelerator}
\newacronym{isol}{ISOL}{isotope separation online}
\newacronym{isosim}{IsoSiM}{Isotopes for Science and Medicine}

\newacronym{lcao}{LCAO}{linear combination of atomic orbitals}
\newacronym{lda}{LDA}{local density approximation}
\newacronym{led}{LED}{light-emitting diode}
\newacronym{leis}{LEIS}{low-energy ion scattering}
\newacronym{lib}{LIB}{lithium-ion battery}
\newacronym{lsat}{LSAT}{\ch{(La,Sr)(Al,Ta)O3}}

\newacronym{mas}{MAS}{magic angle spinning}
\newacronym{mpms}{MPMS}{magnetic property measurement system}
\newacronym{mbe}{MBE}{molecular beam epitaxy}
\newacronym{md}{MD}{molecular dynamics}
\newacronym{midas}{MIDAS}{Maximum Integrated Data Acquisition System}
\newacronym{mit}{MIT}{metal-insulator transition}
\newacronym{mnr}{MNR}{Meyer-Neldel rule}
\newacronym{mqt}{mqt}{multi-quantum transition}
\newacronym{mud}{MUD}{muon data}
\newacronym{ms}{MS}{mass spectrometry}

\newacronym{nbm}{NBM}{neutral beam monitor}
\newacronym{neb}{NEB}{nudged elastic band}
\newacronym{nim}{NIM}{nuclear instrumentation module}
\newacronym{nmr}{NMR}{nuclear magnetic resonance}
\newacronym{no}{NO}{nuclear orientation}
\newacronym{nqr}{NQR}{nuclear quadrupole resonance}
\newacronym{nrc}{NRC}{National Research Council of Canada}
\newacronym{nserc}{NSERC}{Natural Sciences and Engineering Research Council of Canada}

\newacronym{oa}{OA}{optical absorption}

\newacronym{pac}{PAC}{perturbed angular correlation}
\newacronym{pad}{PAD}{perturbed angular distribution}
\newacronym{pas}{PAS}{principle axis system}
\newacronym{pchip}{PCHIP}{piecewise cubic Hermite interpolating polynomial}
\newacronym{pdf}{PDF}{probability density function}
\newacronym{pld}{PLD}{pulsed laser deposition}
\newacronym{ppms}{PPMS}{physical property measurement system}

\newacronym{qens}{QENS}{quasielastic neutron scattering}
\newacronym{ql}{QL}{quintuple layer}
\newacronym{qo}{QO}{quantum oscillations}

\newacronym{rbs}{RBS}{Rutherford backscattering}
\newacronym{rf}{RF}{radio frequency}
\newacronym{rheed}{RHEED}{reflection high-energy electron diffraction}
\newacronym{rib}{RIB}{radioactive ion beam}
\newacronym{rkky}{RKKY}{Ruderman–Kittel–Kasuya–Yosida}
\newacronym{rrr}{RRR}{residual-resistivity ratio}
\newacronym{rtil}{RTIL}{room temperature ionic liquid}

\newacronym{sae}{SAE}{spin-alignment echo}
\newacronym{sans}{SANS}{small angle neutron scattering}
\newacronym{si}{SI}{International System of Units}
\newacronym{sims}{SIMS}{secondary ion mass spectrometry}
\newacronym{slr}{SLR}{spin-lattice relaxation}
\newacronym[sort={S/N}]{snr}{\textit{S}/\textit{N}}{signal-to-noise ratio}
\newacronym{squid}{SQUID}{superconducting quantum interference device}
\newacronym{srf}{SRF}{superconducting radio frequency}
\newacronym{srim}{SRIM}{Stopping and Range of Ions in Matter}
\newacronym{ssid}{SSID}{solid-state ionic device}
\newacronym{ssr}{SSR}{spin-spin relaxation}
\newacronym{stm}{STM}{scanning tunnelling microscopy}
\newacronym{sts}{STS}{scanning tunnelling spectroscopy}

\newacronym{ti}{TI}{topological insulator}
\newacronym{trim}{TRIM}{Transport and Range of Ions in Matter}
\newacronym{tss}{TSS}{topological surface state}
\newacronym{tmd}{TMD}{transition metal dichalcogenide}

\newacronym{uhv}{UHV}{ultra-high vacuum}

\newacronym{vdw}{vdW}{van der Waals}
\newacronym{vft}{VFT}{Vogel-Fulcher-Tammann}

\newacronym{xrd}{XRD}{x-ray diffraction}
\newacronym{xrr}{XRR}{x-ray reflection}

\newacronym{ybco}{YBCO}{\ch{YBa2Cu3O_{6+x}}}
\newacronym{ysz}{YSZ}{yttria-stabilized zirconia}

\newacronym[sort={muSR}]{musr}{\ensuremath{\mu}SR}{muon spin spectroscopy}
\newacronym{alc-musr}{ALC-\ensuremath{\mu}SR}{avoided level crossing muon spin rotation}
\newacronym{le-musr}{LE-\ensuremath{\mu}SR}{low-energy muon spin rotation}
\newacronym{lf-musr}{LF-\ensuremath{\mu}SR}{longitudinal field muon spin rotation}
\newacronym{rf-musr}{RF-\ensuremath{\mu}SR}{radio frequency muon spin rotation}
\newacronym{tf-musr}{TF-\ensuremath{\mu}SR}{transverse field muon spin rotation}
\newacronym{zf-musr}{ZF-\ensuremath{\mu}SR}{zero field muon spin rotation}

\begin{document}

\title{
	Depth-resolved measurement of the Meissner screening profile in a niobium thin film
	from spin-lattice relaxation of the implanted $\beta$-emitter \ch{^{8}Li}
}


\author{Ryan~M.~L.~McFadden}
\email[E-mail: ]{rmlm@triumf.ca}
\affiliation{TRIUMF, 4004 Wesbrook Mall, Vancouver, BC V6T~2A3, Canada}
\affiliation{Department of Physics and Astronomy, University of Victoria, 3800 Finnerty Road, Victoria, BC V8P~5C2, Canada}

\author{Md~Asaduzzaman}
\affiliation{TRIUMF, 4004 Wesbrook Mall, Vancouver, BC V6T~2A3, Canada}
\affiliation{Department of Physics and Astronomy, University of Victoria, 3800 Finnerty Road, Victoria, BC V8P~5C2, Canada}

\author{Terry~J.~Buck}
\altaffiliation[Current address: ]{JSI Telecom, 99 Michael Cowpland Dr, Kanata, ON K2M~1X3, Canada}
\affiliation{Department of Physics and Astronomy, 6224 Agricultural Road, University of British Columbia, Vancouver, BC V6T~1Z1, Canada}

\author{David~L.~Cortie}
\altaffiliation[Current address: ]{Australia's Nuclear Science and Technology Organisation, New Illawarra Road, Lucas Heights, NSW 2234, Australia}
\affiliation{Department of Physics and Astronomy, 6224 Agricultural Road, University of British Columbia, Vancouver, BC V6T~1Z1, Canada}
\affiliation{Department of Chemistry, 2036 Main Mall, University of British Columbia, Vancouver, BC V6T~1Z1, Canada}
\affiliation{Stewart Blusson Quantum Matter Institute, 2355 East Mall, University of British Columbia, Vancouver, BC V6T~1Z4, Canada}

\author{Martin~H.~Dehn}
\altaffiliation[Current address: ]{D-Wave Systems Inc., 3033 Beta Avenue, Burnaby, BC V5G~4M9, Canada}
\affiliation{Department of Physics and Astronomy, 6224 Agricultural Road, University of British Columbia, Vancouver, BC V6T~1Z1, Canada}
\affiliation{Stewart Blusson Quantum Matter Institute, 2355 East Mall, University of British Columbia, Vancouver, BC V6T~1Z4, Canada}

\author{Sarah~R.~Dunsiger}
\affiliation{TRIUMF, 4004 Wesbrook Mall, Vancouver, BC V6T~2A3, Canada}
\affiliation{Department of Physics, Simon Fraser University, 8888 University Drive, Burnaby, BC V5A~1S6, Canada}

\author{Robert~F.~Kiefl}
\affiliation{TRIUMF, 4004 Wesbrook Mall, Vancouver, BC V6T~2A3, Canada}
\affiliation{Department of Physics and Astronomy, 6224 Agricultural Road, University of British Columbia, Vancouver, BC V6T~1Z1, Canada}
\affiliation{Stewart Blusson Quantum Matter Institute, 2355 East Mall, University of British Columbia, Vancouver, BC V6T~1Z4, Canada}

\author{Robert~E.~Laxdal}
\affiliation{TRIUMF, 4004 Wesbrook Mall, Vancouver, BC V6T~2A3, Canada}
\affiliation{Department of Physics and Astronomy, University of Victoria, 3800 Finnerty Road, Victoria, BC V8P~5C2, Canada}

\author{C.~D.~Philip~Levy}
\affiliation{TRIUMF, 4004 Wesbrook Mall, Vancouver, BC V6T~2A3, Canada}

\author{W.~Andrew~MacFarlane}
\affiliation{TRIUMF, 4004 Wesbrook Mall, Vancouver, BC V6T~2A3, Canada}
\affiliation{Department of Chemistry, 2036 Main Mall, University of British Columbia, Vancouver, BC V6T~1Z1, Canada}
\affiliation{Stewart Blusson Quantum Matter Institute, 2355 East Mall, University of British Columbia, Vancouver, BC V6T~1Z4, Canada}

\author{Gerald~D.~Morris}
\affiliation{TRIUMF, 4004 Wesbrook Mall, Vancouver, BC V6T~2A3, Canada}

\author{Matthew~R.~Pearson}
\altaffiliation[Current address: ]{Angstrom Vision Inc., 2-831 3rd St W, North Vancouver, BC V7P~3K7, Canada}
\affiliation{TRIUMF, 4004 Wesbrook Mall, Vancouver, BC V6T~2A3, Canada}

\author{Edward~Thoeng}
\affiliation{TRIUMF, 4004 Wesbrook Mall, Vancouver, BC V6T~2A3, Canada}
\affiliation{Department of Physics and Astronomy, 6224 Agricultural Road, University of British Columbia, Vancouver, BC V6T~1Z1, Canada}

\author{Tobias~Junginger}
\email[E-mail: ]{junginger@uvic.ca}
\affiliation{TRIUMF, 4004 Wesbrook Mall, Vancouver, BC V6T~2A3, Canada}
\affiliation{Department of Physics and Astronomy, University of Victoria, 3800 Finnerty Road, Victoria, BC V8P~5C2, Canada}

\date{\today}

\begin{abstract}
	We report measurements of the Meissner screening profile  
	in a \ch{Nb}(\SI{300}{\nano\meter})/\ch{Al2O3} thin film using \ch{^{8}Li} \gls{bnmr}.
	The \acrshort{nmr} probe \ch{^{8}Li} was ion-implanted into the \ch{Nb} film at energies \SI{\leq 20}{\kilo\electronvolt},
	corresponding to mean stopping depths comparable to \ch{Nb}'s magnetic penetration depth $\lambda$.
	\ch{^{8}Li}'s strong dipole-dipole coupling with the host \ch{^{93}Nb}
	nuclei provided a ``cross-relaxation'' channel that dominated in low magnetic fields,
	which conferred indirect sensitivity to the local magnetic field via the \gls{slr} rate $1/T_{1}$.
	From a fit of the $1/T_{1}$ data to a model accounting for its dependence on temperature, magnetic field,
	and \ch{^{8}Li^{+}} implantation energy,
	we obtained a magnetic penetration depth $\lambda_{0} = \SI{51.5 \pm 2.2}{\nano\meter}$,
	consistent with a relatively short carrier mean-free-path $\ell = \SI{18.7 \pm 2.9}{\nano\meter}$
	typical of similarly prepared \ch{Nb} films.
	The results presented here constitute an important step towards using \ch{^{8}Li} \gls{bnmr} to characterize
	bulk \ch{Nb} samples with engineered surfaces,
	which are often used in the fabrication of particle accelerators.
\end{abstract}

\maketitle
\glsresetall

\section{
	Introduction
	\label{sec:introduction}
}

The transition metal \ch{Nb} is a type-II superconductor with a transition
temperature $T_{c} \approx \SI{9.25}{\kelvin}$~\cite{1966-Finnemore-PR-149-231}
that is highest among the pure elements.
Accompanying this accolade is the highest lower critical field
$B_{c1} \approx \SI{170}{\milli\tesla}$~\cite{1966-Finnemore-PR-149-231} of
\emph{any} known superconductor,
making \ch{Nb} uniquely suited for applications where superconductivity is
required in the presence of (relatively) large magnetic fields.
One such example is its use in \gls{srf} cavities~\cite{2008-Padamsee-RFSA-2,2009-Padamsee-RFSSTA},
which make up many accelerator beamlines (for charged particles) across the globe.
In the pursuit of optimal accelerating efficiency and cryogenic operation economy,
there is great interest in \emph{engineering} \ch{Nb} to remain in the Meissner state
up to the highest magnetic fields possible.
This can be achieved empirically by chemically modifying its surface through
baking~\cite{2004-Ciovati-JAP-96-1591,arXiv:1806.09824}
or
doping~\cite{2013-Grassellino-SST-26-102001,2017-Grassellino-SST-30-094004},
as well as by
coating it with thin layers of other superconducting
(and insulating)
materials~\cite{2006-Gurevich-APL-88-012511,2014-Kubo-APL-104-032603,2015-Gurevich-AIPA-5-017112,2015-Posen-PRA-4-044019,2017-Kubo-SST-30-023001,2019-Kubo-JJAP-58-088001}.

To better understand how these microscopic modifications affect \ch{Nb}'s superconducting properties,
an experimental technique capable of interrogating the \emph{local} electromagnetic fields
in the near surface region is essential.
Similarly, as the material properties are expected to vary below the surface,
\emph{spatial} resolution is also a requirement.
In reality, there are few experimental techniques capable of simultaneously satisfying these criteria.
Perhaps the most widely known is \gls{le-musr}~\cite{2004-Bakule-CP-45-203,2022-Hillier-NRMP-2-4},
which uses implanted positive muons $\mu^{+}$ as local magnetic probes of the field inside .
Indeed,
\gls{le-musr} has been used to study Meissner screening in \ch{Nb},
revealing:
the importance of non-local electrodynamics~\cite{1953-Pippard-PRSLA-216-547,1957-Bardeen-PR-108-1175}
and strong-coupling corrections~\cite{1967-Nam-PR-156-470}
in ``clean'' samples~\cite{2005-Suter-PRB-72-024506};
the dependence of the magnetic penetration depth on film preparation technique~\cite{2017-Junginger-SST-30-125013};
as well as an ``anomalous'' modification to the screening profile for select surface treatments~\cite{2014-Romanenko-APL-104-072601}.
As the latter result is controversial~\cite{2023-McFadden-PRA-19-044018,arXiv:2305.02129},
it is imperative that alternative means of study be developed to corroborate such findings.
To this end,
here we employed a lesser-known (but closely related) technique
--- ion-implanted \gls{bnmr} ---
to study the superconducting properties of \ch{Nb} below its surface.
Thorough accounts of the \gls{bnmr} technique can be found in several
recent~\cite{2015-MacFarlane-SSNMR-68-1, 2022-MacFarlane-ZPC-236-757}
and
older~\cite{1978-Ackermann-ANQR-3-1,1983-Ackermann-TCP-31-291,1995-Widdra-RSI-66-2465,2001-Asahi-NPA-693-63,2005-Heitjans-DCM-9-367,2006-Abov-PAN-69-1701}
review articles, with additional details given in several
technical overviews~\cite{2014-Levy-HI-225-165, 2014-Morris-HI-225-173, 2018-Kreitzman-JPSCP-21-011056}.

In ion-implanted \gls{bnmr}~\cite{2015-MacFarlane-SSNMR-68-1, 2022-MacFarlane-ZPC-236-757},
a radioactive spin probe is ``inserted'' into a
material of interest and its \gls{nmr} response is monitored through the
anisotropic property of $\beta$-decay
(see e.g.,~\cite{1965-Lee-ARNS-15-381}),
analogous to the detection scheme used in \gls{le-musr}~\cite{2004-Bakule-CP-45-203,2004-Morenzoni-JPCM-16-S4583,2022-Hillier-NRMP-2-4}.
Just as in ``conventional'' \gls{nmr} using stable nuclei
(see e.g.,~\cite{1961-Abragam-PNM,1983-Mehring-PHRNMRS,1990-Slichter-PMR,2019-Pell-PNMRS-111-1}),
\gls{bnmr} is sensitive to the electronic and magnetic properties of the ``host'' material;
however, its sensitivity (per nucleus) is greater by a factor of \num{\sim e10},
allowing for the study of situations typically inaccessible by \gls{nmr}
(e.g.,
thin films~\cite{2007-Parolin-PRL-98-047601,2011-Song-PRB-84-054414,2013-MacFarlane-PRB-88-144424,2019-Karner-PRB-100-165109},
heterostructures~\cite{2008-Keeler-PRB-77-144429,2012-Salman-PRL-109-257207,2015-MacFarlane-PRB-92-064409,2021-Karner-PRB-104-205114},
or
isolated impurities~\cite{2014-McKenzie-JACS-136-7833,2017-McFadden-CM-29-10187}).
Supplementing this feature is the ability to control the implantation energy
that the probe radioisotope is ion-implanted,
allowing for it to be ``placed'' at specific depth ranges.
This further extends the types of problems that can be studied,
especially those where \emph{depth-resolution} is essential.
For example,
the depth-resolution afforded by \ch{^{8}Li} \gls{bnmr} has been used to investigate:
molecular dynamics in glassy polystyrene~\cite{2015-McKenzie-SM-11-1755,2018-McKenzie-SM-14-7324,2022-McKenzie-JCP-156-084903};
the magnetism of \ch{Mn12} single-molecule magnets~\cite{2007-Salman-NL-7-1551};
surface magnons in antiferromagnetic \ch{Fe2O3}~\cite{2016-Cortie-PRL-116-106103};
finite size effects in \ch{Pd} thin films~\cite{2013-MacFarlane-PRB-88-144424};
as well as the superconductor \ch{NbSe2} in both
the vortex~\cite{2007-Salman-PRL-98-167001}
and
Meissner~\cite{2009-Hossain-PRB-79-144518} states.

While \gls{le-musr} and \gls{bnmr} are largely complementary techniques~\cite{2000-Kiefl-PB-289-640},
differences in probe properties
(e.g., radioactive lifetime)
make \gls{bnmr} much more amenable to the measurement of phenomena transpiring over longer timescales,
such as \gls{slr}
(in non-magnetic materials).
\Gls{slr}
is driven by stochastic fluctuations in the local electromagnetic field at the probe's stopping site,
which dynamically ``re-orient'' the probe's spin,
returning the polarization of the ensemble to \num{\sim 0}
(i.e., its thermal equilibrium value).
When the ``host'' contains an abundance of isotopes with non-zero nuclear spin,
the implanted probe couples to the ``bath'' of host spins via the dipolar interaction,
providing a relaxation ``channel'' with sensitivity to the static component of the local field.
This form of ``cross-relaxation''
(see e.g.,~\cite{2012-Chow-PRB-85-092103})
has been used to study the
magnetic penetration depth in the multi-band superconductor \ch{NbSe2}~\cite{2009-Hossain-PRB-79-144518},
as well as order-parameter fluctuations in superconducting \ch{Pb} and \ch{Ag/Nb} thin films~\cite{2012-Morenzoni-PRB-85-220501},
by way of the \gls{bnmr} probe \ch{^{8}Li}.
Here,
we applied and extended this methodology to extract the Meissner screening
profile in a \ch{Nb}(\SI{300}{\nano\meter})/\ch{Al2O3} thin film.
Specifically,
in a \SI{20}{\milli\tesla} applied magnetic field, 
we measured the temperature and implantation energy (i.e., depth) dependence
of implanted \ch{^{8}Li}'s \gls{slr} rate $1/T_{1}$,
revealing a sharp increase in relaxation at temperatures below $T_{c}$,
whose magnitude increases deeper below the material's surface.
By contrast,
this depth-dependence vanishes in the normal state,
consistent with the expected signature for Meissner screening~\cite{2009-Hossain-PRB-79-144518,2012-Morenzoni-PRB-85-220501}.
From a fit of the data to a model accounting for $1/T_{1}$'s
dependence on temperature, implantation energy, and magnetic field,
we determined that the screening profile is well-described by a
simple London model~\cite{1935-London-PRSLA-149-71,1996-Tinkham-IS-2}
with a magnetic penetration depth
exceeding \ch{Nb}'s ``intrinsic'' value.
The result is consistent with similarly prepared films,
known to possess relatively short carrier mean-free-paths.

The rest of the manuscript is organized as follows.
Details of the \gls{bnmr} experiments are given in \Cref{sec:experiment},
along with information on the preparation of our \ch{Nb} sample
(\Cref{sec:experiment:film}).
Our results and analysis are presented in \Cref{sec:results},
with a description of our data given in \Cref{sec:results:data},
followed by an account of our analysis approach (\Cref{sec:results:rates}),
and the modelling of the extracted results (\Cref{sec:results:model}).
We discuss our findings and their implications in \Cref{sec:discussion},
and a summary is given in \Cref{sec:conclusions}.

\section{
	Experiment
	\label{sec:experiment}
}

\gls{bnmr} experiments were performed at
TRIUMF's \gls{isac} facility in Vancouver, BC, Canada.
The probe radioisotope \ch{^{8}Li}
(nuclear spin $I = 2$;
gyromagnetic ratio
$\gamma / (2 \pi) = \SI{6.30221 \pm 0.00007}{\mega\hertz\per\tesla}$~\cite{2019-Stone-INDC-NDS-0794,2021-Tiesinga-RMP-93-025010};
electric quadrupole moment
$Q = \SI[retain-explicit-plus=true]{+31.4 \pm 0.2}{\milli b}$~\cite{2021-Stone-INDC-NDS-0833};
radioactive lifetime
$\tau_{\beta} = \SI{1.2096 \pm 0.0005}{\second}$~\cite{2010-Flechard-PRC-82-027309};
and mass 
$m_{\ch{^{8}Li}} = \SI{8.022 486 24 \pm 0.000 000 05}{\amu}$~\cite{2021-Huang-CPC-45-030002,2021-Wang-CPC-45-030003})
was generated from an isotope production target
(e.g., \ch{Ta} foil stacks)~\cite{2014-Bricault-HI-225-25}
irradiated with a \SI{\sim 500}{\mega\electronvolt} proton beam at
a current on the order of \SI{\sim 20}{\micro\ampere}.
The target, heated in excess of \SI{\sim 2000}{\kelvin} to accelerate thermal
out-diffusion of \ch{^{8}Li},
was coupled to a surface ion source~\cite{2014-Bricault-HI-225-25}
and
a mono-energetic \SI{\sim 20}{\kilo\electronvolt} beam of \ch{^{8}Li^{+}}
was extracted with an intensity of \SI{\sim e7}{ions\per\second}.
A high-resolution magnetic dipole mass separator was used to ensure the
isotopic purity of the beam,
which was transported through a high vacuum beamline via electrostatic optics.
During transport, the ion beam was neutralized in-flight and subsequently
spin-polarized by collinear optical pumping with circularly polarized resonant
laser light~\cite{2014-Levy-HI-225-165},
yielding a high degree of nuclear spin-polarization
(\SI{\sim 70}{\percent}~\cite{2003-Levy-NIMB-204-689,2014-MacFarlane-JPCS-551-012059}).
The highly polarized beam was then re-ionized and (electrostatically) steered
to the spectrometer~\cite{2014-Morris-HI-225-173,2015-MacFarlane-SSNMR-68-1},
where the probe \ch{^{8}Li^{+}} ions were implanted
into the \ch{Nb} thin film,
which was affixed to a holder compatible with the spectrometer's cold-finger cryostat.
The energy of the ion beam was controlled by biasing the spectrometer's
platform to high positive voltage (up to \SI{20}{\kilo\volt}),
electrostatically decelerating the \ch{^{8}Li^{+}} from their transport energy
(within the final few centimeters of their trajectory)
before implantation.
Note that, in the lead-up to implantation,
the \ch{^{8}Li} spin-polarization is parallel to the film's
surface and the applied field,
but perpendicular to the beam's momentum
(see e.g.,~\cite{2014-Morris-HI-225-173}).

\begin{figure*}
	\centering
	\includegraphics[width=1.0\textwidth]{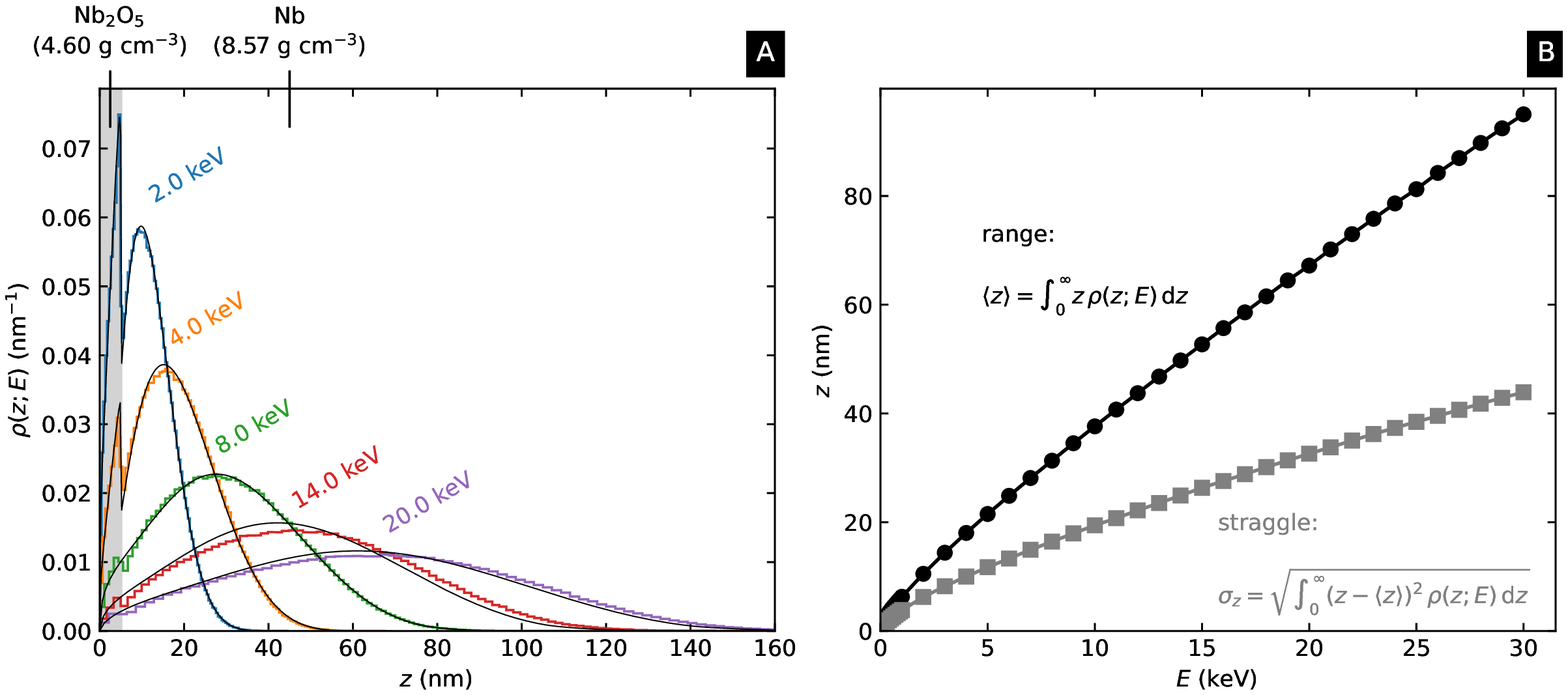}
	\caption{
		\label{fig:srim}
		Results from simulations of \num{e6} \ch{^{8}Li^{+}} ions implanted in a
		\ch{Nb2O5}(\SI{5}{\nano\meter})/\ch{Nb} target,
		calculated using the \gls{srim} Monte Carlo code~\cite{srim,2010-Ziegler-NIMB-268-1818}
		by aid of the Python package pysrim~\cite{2018-Ostrouchov-JOSS-3-829}.
		(\textbf{A}):
		Typical \ch{^{8}Li^{+}} stopping profiles at select implantation energies
		$E$ are shown, represented here as histograms (with \num{100} bins)
		of the stopping probability $\rho(z;E)$ vs.\ depth below the surface $z$.
		The thin black lines denote fits to a phenomenological model for $\rho(z;E)$
		[\Cref{eq:stopping,eq:beta-pdf} --- see \Cref{sec:results:model}],
		capturing all features of the stopping profiles at each energy.
		Note that the abrupt change in $\rho(z;E)$ at the \ch{Nb2O5}/\ch{Nb}
		boundary is related to the different layer densities
		(indicated in the inset).
		The thickness of the surface pentoxide oxide layer (\SI{5}{\nano\meter})
		is typical for \ch{Nb} (see e.g.,~\cite{1987-Halbritter-APA-43-1}).
		(\textbf{B}):
		Moments of the stopping profiles.
		As $E$ increases, so does the
		\ch{^{8}Li^{+}} range and straggle
		(i.e., the mean, $\langle z \rangle$, and standard deviation, $\sigma_{z}$).
	}
\end{figure*}

An essential aspect of this approach is the spatial (i.e., depth) resolution
obtained through ion-implantating our \gls{bnmr} probe \ch{^{8}Li} into \ch{Nb}.
Ion-implantation is a controlled, non-equilibrium means of introducing
impurities in solids, which finds use (on an industrial scale), for example,
in the electrical doping of semiconductors (see e.g.,~\cite{2006-Nastasi-IISM}).
During implantation, the ``injected'' ion loses energy through its
interaction with the host's electrons and nuclei,
amounting to a series of (dampened) collisions that thermalize the projectile.
The slowing process is stochastic,
resulting in a \emph{distribution} of stopping positions for a mono-energetic
projectile.
This process can be simulated accurately using Monte Carlo methods
(see e.g.,~\cite{1991-Eckstein-SSMS-10,srim}),
and the results from such simulations are used to connect the probe's \gls{nmr}
response with a spatial region of the sample,
just as in \gls{le-musr}~\cite{2002-Morenzoni-NIMB-192-245,2020-Somoes-RSI-91-023906}.
Several implantation profiles for
\SIrange{2}{20}{\kilo\electronvolt} \ch{^{8}Li^{+}}
stopping in \ch{Nb},
simulated using the \gls{srim} Monte Carlo code~\cite{srim,2010-Ziegler-NIMB-268-1818},
are shown in \Cref{fig:srim}.

\subsection{
	Measurements
	\label{sec:experiment:bnmr}
}

In each \gls{bnmr} measurement,
the \ch{^{8}Li} spin-polarization was monitored after implantation
through the anisotropy of its radioactive $\beta$-decay~\cite{1965-Lee-ARNS-15-381}.
Specifically, the experimental asymmetry $A$
(proportional to the average longitudinal spin-polarization $p_{z}$)
was measured by combining the $\beta$-rates in
two opposed scintillation counters~\cite{2015-MacFarlane-SSNMR-68-1}.
In our measurements,
the sense of circular polarization (left and right)
of the pumping laser light was also alternated~\cite{2014-Levy-HI-225-165},
producing either ``positive'' or ``negative'' helicity in the \ch{^{8}Li^{+}} beam
(i.e., its nuclear-spin polarization is aligned or counter-aligned with the beam direction).
Data were collected separately for each helicity and then combined to remove
detection systematics
(see e.g.,~\cite{1995-Widdra-RSI-66-2465, 2015-MacFarlane-SSNMR-68-1}).
Explicitly, the well-known ``four-counter'' method was used:
\begin{equation} \label{eq:asymmetry}
   A = \frac{1 - r}{1 + r} \equiv A_{0} p_{z},
\end{equation}
where
\begin{equation*}
   r \equiv \sqrt{ \frac{ \left ( L_{+} / R_{+} \right ) }{ \left ( L_{-} / R_{-} \right ) } }
\end{equation*}
is the geometric mean of the ratio of rates in the opposing counters
$L$ and $R$ for the $\pm$ polarization senses.
Note that the proportionality constant between $A$ and $p_{z}$,
denoted as $A_{0}$ in \Cref{eq:asymmetry},
depends on both the experimental geometry and
the details of the $\beta$-decay
(on the order of \SI{\sim 10}{\percent} here).

\Gls{slr} measurements were performed by monitoring
the transient decay of spin-polarization both during and
following a short pulse of beam of duration $\Delta$
(typically \SI{\sim 4}{\second}).
During the pulse, the polarization approaches a steady-state value,
while after the pulse, it relaxes to essentially zero.
At the edge of the pulse, there is a discontinuity in the slope,
characteristic of \gls{bnmr} \gls{slr} data
acquired in this manner~\cite{2015-MacFarlane-SSNMR-68-1,2022-MacFarlane-ZPC-236-757}.
This bipartite behavior can be described quantitatively by considering a
uniform distribution of probe arrival times
(i.e., a time-independent \ch{^{8}Li^{+}} production rate). 
Explicitly, for an arrival time $t^{\prime}$, the temporal evolution of
nuclear spin-polarization at time $t > t^{\prime}$ follows: 
\begin{widetext}
\begin{equation}
	\label{eq:asymmetry-time}
   p_{z}(t) = p_{0} \times
   \begin{cases}
         \dfrac{ \displaystyle \int_{0}^{t} \exp \left[ - \left ( t - t^{\prime} \right ) / \tau_{\beta} \right ] R(t, t^{\prime}) \, \mathrm{d}t^{\prime} }{ \displaystyle \int_{0}^{t} \exp \left[ - t^{\prime} / \tau_{\beta} \right ] \, \mathrm{d}t^{\prime} } , & t \leq \Delta , \\
         \dfrac{ \displaystyle \int_{0}^{\Delta} \exp \left[ - \left ( t - t^{\prime} \right ) / \tau_{\beta} \right ] R(t, t^{\prime}) \, \mathrm{d}t^{\prime} }{ \exp \left[ - \left ( t - \Delta \right ) / \tau_{\beta} \right ] \displaystyle \int_{0}^{\Delta} \exp \left[ - t^{\prime} / \tau_{\beta} \right ] \, \mathrm{d}t^{\prime} } , & t > \Delta ,
      \end{cases}
\end{equation}
\end{widetext}
where $p_{0} \approx \SI{70}{\percent}$ is the degree of nuclear spin-polarization
produced from optical pumping~\cite{2003-Levy-NIMB-204-689,2014-MacFarlane-JPCS-551-012059},
and
$R(t, t^{\prime})$ is a ``relaxation'' function describing the decay of $p_{z}$
(e.g., an exponential in the simplest case).
Note that,
unlike in conventional \gls{nmr},
no transverse \gls{rf} field $B_{1}$ is required for measurement of the \gls{slr},
meaning that there is no intrinsic spectral resolution of the relaxation,
which represents the \gls{slr} of the \emph{entire} \ch{^{8}Li} population
(i.e., not just those with observable resonances).
A typical \gls{slr} measurement required \SI{\sim 30}{\minute},
corresponding to \num{\sim e8} total decay events.

\subsection{
	Sample
	\label{sec:experiment:film}
}

A \SI{300}{\nano\meter} \ch{Nb} thin film from a previous \gls{bnmr}
experiment~\cite{2009-Parolin-PRB-80-174109} was employed in this study.
The film was grown (\SI{1}{\angstrom\per\second} deposition rate)
by \gls{rf} sputtering a \SI{99.9}{\percent} \ch{Nb} target (Goodfellow)
onto an 
\SI{8 x 12 x 0.5}{\milli\meter}
epitaxially polished \ch{Al2O3} $\langle 0001 \rangle$
substrate (Honeywell) at \SI{\sim 50}{\celsius} in \SI{3}{\milli Torr} of
\ch{Ar} (\SI{25}{\centi\meter\cubed\per\minute} flow),
resulting in a high degree of orientation in the $\langle 110 \rangle$ direction
(confirmed by \gls{xrd}).
The film had a bulk critical temperature $T_{c} \approx \SI{8.9}{\kelvin}$,
as determined from magnetometry measurements.
Further characterization details are given
elsewhere~\cite{2009-Parolin-PRB-80-174109}.

\section{
	Results \& Analysis
	\label{sec:results}
}

This section is divided into three parts.
First, we present the raw \gls{slr} data and describe its features in
\Cref{sec:results:data}.
Subsequently, we consider the analysis of the \gls{slr} data and show the
extracted \gls{slr} rates ($1/T_{1}$) in \Cref{sec:results:rates}.
Finally, in \Cref{sec:results:model}, we construct a model for $1/T_{1}$ that
allows for the extraction of the magnetic penetration depth
and apply it to the results.

\subsection{
	\acrshort{slr} Data
	\label{sec:results:data}
}

\begin{figure*}
	\centering
	\includegraphics[width=1.0\textwidth]{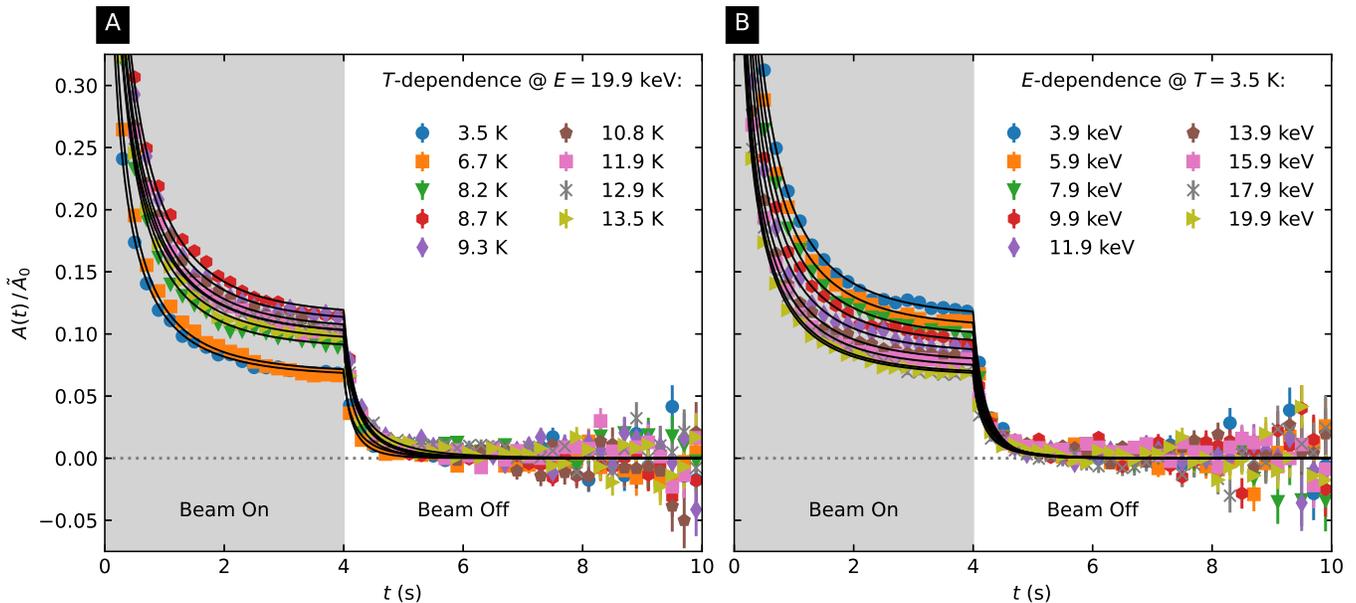}
	\caption{
		\label{fig:slr-data}
		Typical \ch{^{8}Li} \gls{slr} data in \ch{Nb}(\SI{300}{\nano\meter})/\ch{Al2O3}
		at low temperatures $T < \SI{15}{\kelvin}$
		and implantation energies $E \leq \SI{20}{\kilo\electronvolt}$
		(where all implanted \ch{^{8}Li^{+}} stops in the film --- see \Cref{fig:srim})
		in an applied magnetic field
		$B_{0} = \SI{20}{\milli\tesla}$ parallel to the sample surface.
		The data (represented as histograms) have been binned by a factor of
		\num{20} for clarity and the solid lines represent fits using a stretched
		exponential relaxation function
		(described in \Cref{sec:results:rates}).
		In all cases, the \gls{slr} rate is orders of magnitude faster than
		at higher fields~\cite{2009-Parolin-PRB-80-174109}
		with much of the (normalized) asymmetry $A(t) / \tilde{A}_{0}$
		vanishing within the first few milliseconds
		(the vertical scale has been adjusted to better show the remaining signal). 
		(\textbf{A}):
		Temperature dependent measurements at a constant implantation
		energy $E = \SI{19.9}{\kilo\electronvolt}$.
		Upon cooling the sample from the normal state ($T > T_{c}$) to the
		Meissner state ($T < T_{c}$) the \gls{slr} rate increases,
		as evidenced by the increased haste in the decay of the ``curves''.
		(\textbf{B}):
		Implantation energy dependent measurements at a constant temperature
		$T = \SI{3.5}{\kelvin}$.
		In the Meissner state,
		as the implantation energy increases
		(equivalent to \ch{^{8}Li} sampling deeper below \ch{Nb}'s surface)
		the \gls{slr} rate increases.
		Note that equivalent measurements in the normal state ($T > T_{c}$)
		show no such dependence on energy.
	}
\end{figure*}

Typical \ch{^{8}Li} \gls{slr} data in \ch{Nb} at low temperatures and
magnetic fields are shown in \Cref{fig:slr-data}.
In all cases,
the \gls{slr} is orders of magnitude faster than
in fields on the order of \SI{\sim 1}{\tesla}~\cite{2009-Parolin-PRB-80-174109},
typical of measurements in the field regime
where low-frequency fluctuations of the ``host'' \emph{nuclear} spins
provide the dominant relaxation mechanism
(see e.g.,~\cite{2009-Hossain-PRB-79-144518,2018-MacFarlane-JPSCP-21-011020}).
The relaxation,
however,
while fast relative to the ``natural'' timescale of the probe
($\sim 1/ \tau_{\beta}$), 
remains observable for all conditions investigated here.
From the temperature dependent measurements at a constant implantation
energy $E = \SI{19.9}{\kilo\electronvolt}$
(i.e., $\langle z \rangle \approx \SI{70}{\nano\meter}$),
it is apparent that,
upon cooling the sample from the normal state ($T > T_{c}$) to the
Meissner state ($T < T_{c}$),
the \gls{slr} rate increases
(as evidenced by the increased haste in the decay of the \gls{slr} ``curves'').
This \gls{slr} behavior is consistent with that observed in
other superconductors~\cite{2009-Hossain-PRB-79-144518,2012-Morenzoni-PRB-85-220501}.
Similarly,
from measurements at constant temperature $T = \SI{3.5}{\kelvin}$
(well into the Meissner state),
it is clear that as the implantation energy increases
(equivalent to \ch{^{8}Li} sampling deeper below \ch{Nb}'s surface),
so too does the \gls{slr} rate.
Note that equivalent measurements in the normal state ($T > T_{c}$)
show no such dependence on $E$,
suggesting that this \gls{slr} response is a result of \ch{Nb}'s
superconductivity.
Interestingly,
we note that the data in \Cref{fig:slr-data} includes measurements
both where the applied field is zero during cool-down
(zero-field-cooled)
and
when it is \SI{20}{\milli\tesla}
(field-cooled),
showing no appreciable differences
(see below).

\subsection{
	Quantifying $1/T_{1}$
	\label{sec:results:rates}
}

To quantify the observations outlined in \Cref{sec:results:data},
we consider a model to determine the \gls{slr} rate $1/T_{1}$.
First,
we remark that,
in contrast to measurements at higher fields~\cite{2009-Parolin-PRB-80-174109},
the decay of the \gls{slr} ``curves'' is \emph{non-exponential}
at all measured conditions.
This isn't unprecedented for a spin $I = 2$ nucleus,
though it is difficult to identify the exact underlying cause
(e.g.,~\cite{1984-Stockmann-JNCS-66-501,1995-McDowell-JMRSA-113-242,1970-Hubbard-JCP-53-985,1982-Becker-ZNA-37-697}),
which may be complicated by a fundamental change in the character of the \gls{slr}
at low magnetic fields
(see e.g.,~\cite{2018-MacFarlane-JPSCP-21-011020}).
This detail is compounded by the fact that there are \emph{two} \ch{^{8}Li^{+}}
stopping sites in \ch{Nb} at low temperature~\cite{2009-Parolin-PRB-80-174109}.
While an atomistic model of the stopping sites may illuminate the matter,
it is still forthcoming~\cite{Adelman-tbp}.
Consequently,
we elected for a pragmatic approach 
and fit the \gls{slr} data using a (phenomenological) stretched exponential
``relaxation'' function.
For a uniform distribution of probe arrival times
(i.e., a time-independent \ch{^{8}Li^{+}} production rate),
$R(t,t^{\prime})$ from \Cref{eq:asymmetry-time} takes the form: 
\begin{equation}
	\label{eq:strexp}
	R(t, t^{\prime}) = \exp \left [ - \left ( \frac{ t - t^{\prime} }{ T_{1} } \right )^{\beta} \right ] ,
\end{equation}
where $1 / T_{1}$ is the \gls{slr} rate,
and $\beta \in (0, 1]$ is the stretching exponent.
While a value of $\beta < 1$ modifies the form of an exponential decay,
equivalent to the presence of faster (slower) relaxation at early (late) times,
the quantity $T_{1}$ captures the characteristic $1/e$ decay time of $p_{z}$.
We note that this model accurately describes the data over \emph{all}
measurement conditions,
without over-parameterization~\footnote{Based on information about the \ch{^{8}Li} stopping site (inferred from its resonance)~\cite{2009-Parolin-PRB-80-174109}, a more fundamental analysis might use a discrete sum of exponentials for $R(t,t^{\prime})$~\cite{2018-MacFarlane-JPSCP-21-011020}; however, we find the quality of the present data inadequate for such a treatment, which yields ill-defined fit parameters due to over-paramaterization. This phenomenon is not unique to \ch{Nb} and simplified relaxation functions have often been used to quantify the \gls{slr} in other ``simple'' cubic metals (see e.g.,~\cite{2015-MacFarlane-SSNMR-68-1}).}.

\begin{figure*}
	\centering
	\includegraphics[width=1.0\textwidth]{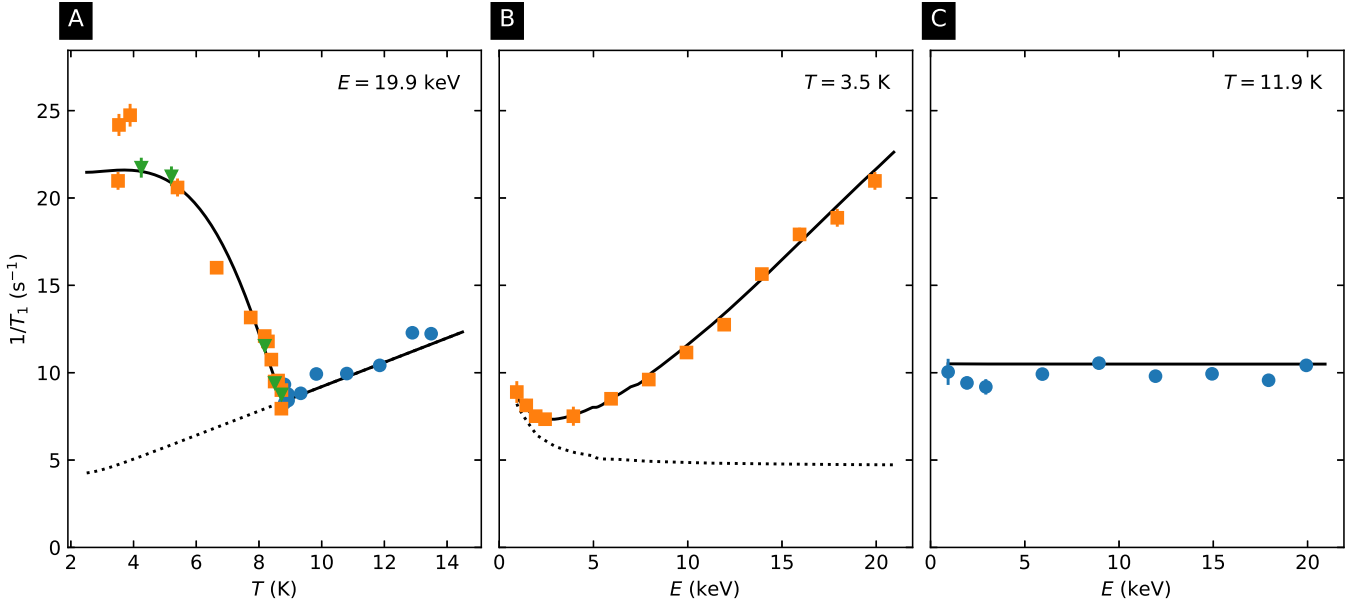}
	\caption{
		\label{fig:slr-rates-film}
		\ch{^{8}Li} \gls{slr} rates $1/T_{1}$ in \ch{Nb}(\SI{300}{\nano\meter})/\ch{Al2O3}
		at \SI{20}{\milli\tesla} parallel to its surface,
		determined from fits to a stretched exponential model
		[\Cref{eq:asymmetry,eq:asymmetry-time,eq:strexp}]
		described in the text.
		The symbols distinguish different groups of data
		[normal state ($\bullet$);
		Meissner state, field-cooled ($\blacksquare$);
		Meissner state, zero-field-cooled ($\blacktriangledown$)],
		all in good agreement with each other.
		The solid black line (---) denotes a fit of the $1/T_{1}(E, T, B)$ model
		to \emph{all} the data, clearly capturing all of the main features.
		For comparison,
		the dotted black line ($\cdot\cdot\cdot$) shows the predicted normal-state
		response for the \gls{slr} rate.
		(\textbf{A})
		Temperature dependence of $1/T_{1}$ at constant implantation energy ($E = \SI{19.9}{\kilo\electronvolt}$),
		corresponding to a mean stopping depth of \SI{\sim 90}{\nano\meter}.
		In the normal state, the \gls{slr} rate varies linearly with temperature,
		but is orders of magnitude larger than the electronic
		(i.e., Korringa~\cite{1950-Korringa-P-16-601,1990-Slichter-PMR})
		relaxation observed at higher applied fields~\cite{2009-Parolin-PRB-80-174109}.
		At $T_{c} \approx \SI{8.8}{\kelvin}$, there is a pronounced kink in the temperature dependence and
		$1/T_{1}$ increases substantially with decreasing $T$.
		This is the manifestation of Meissner screening sensed through the dipole-dipole coupling
		(i.e., ``cross-relaxation'') with the \ch{Nb} host's nuclear spins
		(see e.g.,~\cite{2009-Hossain-PRB-79-144518,2012-Morenzoni-PRB-85-220501}).
  		(\textbf{B})
		Implantation energy dependence of $1/T_{1}$ at constant temperature in the Meissner state ($T = \SI{3.5}{\kelvin}$).
		At energies \SI{> 4}{\kilo\electronvolt}, the \gls{slr} rate increases monotonically with increasing $E$,
		consistent with enhanced field screening deeper below the surface.
		At energies \SI{\leq 4}{\kilo\electronvolt}, there is a small upturn in $1/T_{1}$ with decreasing $E$,
		likely a result of a (weak) paramagnetic contribution from the native oxide layer at the \ch{Nb} surface~\cite{1991-Cava-PRB-44-6973,2011-Proslier-IEEETAS-21-2619}.
		(\textbf{C})
		Implantation energy dependence of $1/T_{1}$ at constant temperature in the normal state ($T = \SI{11.9}{\kelvin}$).
		In contrast to the data in the Meissner state,
		no meaningful energy (i.e., depth) dependence is observed.
	}
\end{figure*}

\begin{figure}
	\centering
	\includegraphics[width=1.0\columnwidth]{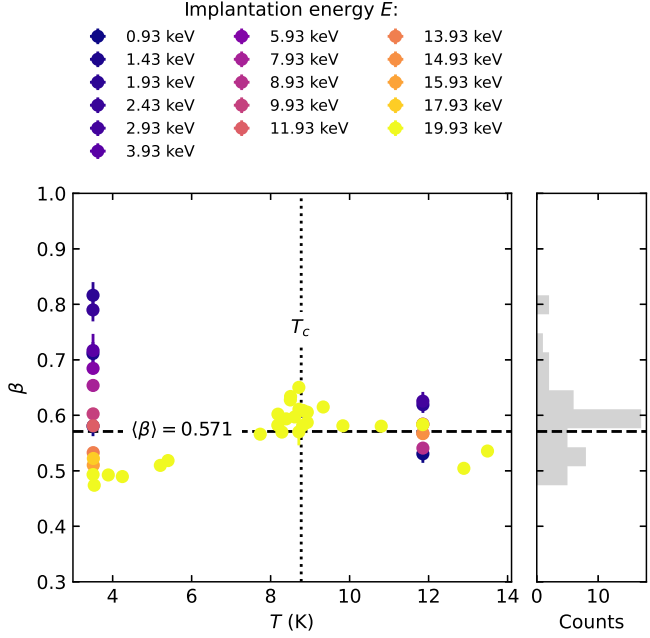}
	\caption{
		\label{fig:beta}
		The temperature ($T$) and implantation energy ($E$) dependence of the
		stretching exponent $\beta$,
		determined from fits of the \ch{^{8}Li} \gls{slr} data (\Cref{fig:slr-data})
		to a stretched exponential relaxation function [\Cref{eq:strexp}].
		$\beta$ is found to be weakly dependent on both $T$ and $E$,
		but remains close to \num{\sim 0.5} for the majority of our measurement conditions.
		Above \ch{Nb}'s critical temperature $T_{c}$
		(indicated by the vertical dotted line),
		$\beta$ is nearly constant and independent of implantation energy $E$,
		showing only a small degree of scatter.
		At $T_{c}$, a small local maximum is observed,
		with the exponent decreasing smoothly as the temperature is lowered.
		At \SI{\sim 3.5}{\kelvin}
		(i.e., well into the Meissner state),
		the $E$-dependence is most pronounced,
		with the (depth-averaged) relaxation becoming increasingly homogeneous
		with decreasing $E$.
		The exponent's average value
		($\langle \beta \rangle \approx 0.571$)
		is indicated by the horizontal dashed line,
		with the value included in the inset.
		The distribution of $\beta$ values is also shown as a histogram.
	}
\end{figure}

The \gls{slr} data were fit using \Cref{eq:asymmetry,eq:asymmetry-time,eq:strexp}.
Note that, for this definition of $A(t)$, the terms $A_{0}$ and $p_{0}$ are not
uniquely defined and instead we used a single ``effective'' parameter,
$\tilde{A}_{0} \equiv A_{0} p_{0}$,
which is on the order of \SI{\sim 10}{\percent} for our data.
Similarly, to mitigate the known ``inflation'' of $\tilde{A}_{0}$ from \Cref{eq:strexp}
(i.e., from the divergence in \Cref{eq:asymmetry-time,eq:strexp} when $t \rightarrow \SI{0}{\second}$),
we constrained our fits such that $\tilde{A}_{0}$ was shared
across measurements at constant $E$
(i.e., in a so-called ``global'' fit).
Empirically,
$\tilde{A}_{0}$ was found to vary with $E$ according to a piecewise linear function
(i.e., with different slopes above and below \SI{\sim 4}{\kilo\electronvolt})
and we imposed the additional constraint that it must fall along this
line~\footnote{We stress that neither of these choices fundamentally altered the observed trends in $1/T_{1}$ (see \Cref{fig:slr-rates-film}).}.
Lastly, we point out that for our $R(t, t^{\prime})$,
numeric integration is required to evaluate \Cref{eq:asymmetry-time}.
For this, we used tanh-sinh quadrature~\cite{1974-Takahashi-PRIMSKU-9-721,2001-Mori-JCAM-127-287},
(i.e., a so-called ``double-exponential'' technique),
which provides both accurate and expedient evaluation of our model.
The fit quality was good for all the \gls{slr} data,
with a typical reduced $\chi^{2} \approx 1.02$ for each measurement.
The fitting was performed by a custom Python script
using the mudpy, bdata, and bfit libraries~\cite{arXiv:2004.10395,2021-Fujimoto-JOSS-6-3598}.
A subset of the fit results are shown in
\Cref{fig:slr-data},
with the \gls{slr} rates and stretching exponents
extracted from the analysis presented in
\Cref{fig:slr-rates-film,fig:beta}.

\begin{table*}
	\caption{
		\label{tab:results}
		Summary of the main fit parameters and their uncertainties
		(rounded according to the Particle Data Group's rules~\cite{2022-Workman-PTEP-2022-083C01-short}),
		determined from fits of the $1/T_{1}(E, B, T)$ model to the data in \Cref{fig:slr-rates-film}
		(solid black line).
		For convenience,
		a brief description of each parameter
		and
		a list of equations where they appear
		are also included.
		Further details about the model and fitting procedure are given in \Cref{sec:results:model}.
	}
	\begin{ruledtabular}
		\begin{tabular}{l l l l l}
		Symbol & Value & Units & Description & Appearances \\
		\hline
		$d$ & $5.0 \pm 0.9$ & nm & thickness of the non-superconducting ``dead layer'' at the film's surface & \Cref{eq:london-film} \\
		$\lambda_{0}$ & $51.5 \pm 2.2$ & nm & the film's magnetic penetration depth & \Cref{eq:london-film,eq:two-fluid} \\
		$B_{d}$ & $(41.1 \pm 2.1) \times 10^{-6}$ & T & magnitude of the dipolar field at the \ch{^{8}Li} stopping sites & \Cref{eq:dipole} \\
		$\tau_{c}$ & $(6.1 \pm 0.5) \times 10^{-6}$ & s & \gls{nmr} correlation time for the fluctuations causing \gls{slr} & \Cref{eq:spectral-density} \\
		$T_{c}$ & $8.775 \pm 0.014$ & K & the film's superconducting transition temperature & \Cref{eq:two-fluid} \\
		$c$ & $0.697 \pm 0.011$ & s$^{-1}$ K$^{-1}$ & slope of the $T$-linear contribution to the total \gls{slr} rate & \Cref{eq:linear} \\
		$C$ & $(2.03 \pm 0.19) \times 10^{3}$ & s$^{-1}$ K$^{b}$ & constant in the near-surface paramagnetic \gls{slr} contribution & \Cref{eq:Nb2O5} \\
		$b$ & $4.5$ &  & exponent in the near-surface surface paramagnetic \gls{slr} contribution & \Cref{eq:Nb2O5} \\
		\end{tabular}
	\end{ruledtabular}
\end{table*}

We now consider the results from fitting.
In the normal state below \SI{\sim 15}{\kelvin},
$1/T_{1}$ varies approximately linearly with temperature
(see \Cref{fig:slr-rates-film});
however,
its slope is much larger
than in measurements at higher field~\cite{2009-Parolin-PRB-80-174109},
resembling the \ch{^{8}Li} \gls{slr} in \ch{Au}~\cite{2018-MacFarlane-JPSCP-21-011020}
and
other elemental metals~\cite{MacFarlane-tbp}.
Similarly,
the (extrapolated) intercept of $1/T_{1}$ at \SI{0}{\kelvin} is also non-zero.
Note that these two features are inconsistent with
the so-called Korringa response~\cite{1950-Korringa-P-16-601,1990-Slichter-PMR}
expected for relaxation due to electrons in a metal's conduction band.
We shall return to this point below.

Just below the film's ``bulk'' $T_{c} \approx \SI{8.9}{\kelvin}$,
there is a pronounced ``kink'' in the temperature dependence of $1/T_{1}$.
The discontinuity at \SI{\sim 8.8}{\kelvin} is suggestive of a transition temperature,
with the shift to lower-$T$ being consistent with the suppression of $T_{c}$
by the small (\SI{20}{\milli\tesla}) applied field.
As the temperature is lowered further,
$1/T_{1}$ increases non-linearly,
with the sharpest increase occurring within the first few Kelvin below the ``kink''.
At colder temperatures,
the increased \gls{slr} ceases,
with $1/T_{1}$ saturating by \SI{\sim 4}{\kelvin}.
This $1/T_{1}$ ``dome'' is consistent with the expected \ch{^{8}Li} \gls{slr}
signature for Meissner screening~\cite{2009-Hossain-PRB-79-144518,2012-Morenzoni-PRB-85-220501}.
Interestingly, as noted earlier,
both field- and zero-field-cooled measurements are in good agreement with each other,
suggesting that either
flux trapping has little influence on the observed \gls{slr} rates
or
there is little flux trapped in the film.

Interestingly,
we find a stretching exponent $\beta$ close to \SI{\sim 0.5} at all measurement conditions,
showing only weak variation with temperature or implantation energy
(see \Cref{fig:beta}).
Above \ch{Nb}'s $T_{c}$,
$\beta$ is nearly constant and independent of $E$,
showing only a small degree of scatter.
Near $T_{c}$, a small local maximum is observed,
with the exponent decreasing smoothly as the temperature is lowered.
At the lowest measured temperature (\SI{\sim 3.5}{\kelvin}),
some $E$-dependence to $\beta$ is evident,
with $\beta$ increasing with decreasing $E$,
implying the \gls{slr} becomes more homogeneous closer to the sample's surface.
While $\beta = 0.5$ is often ascribed to heterogenous relaxation resulting from,
for example, spatial disorder~\cite{1984-Stockmann-JNCS-66-501},
it's absence at high field~\cite{2009-Parolin-PRB-80-174109} suggests it must be of a different origin
(cf.\ the field-dependent \ch{^{8}Li} \gls{slr} in \ch{Au}~\cite{2018-MacFarlane-JPSCP-21-011020}).
Given the ambiguity of isolating the cause the non-exponential \gls{slr} transients,
we suggest that the $\beta$ of \num{\sim 0.5} represents an
effective average over all relaxing \ch{^{8}Li},
whose \gls{slr} time-constants are too similar to isolate here.

Lastly,
we note that
the implantation energy dependence in the Meissner state
(\SI{3.5}{\kelvin})
appears more complicated than anticipated.
As $E$ increases, $1/T_{1}$ decreases slightly,
reaching a minimum around \SI{\sim 2.5}{\kilo\electronvolt}
(i.e., $\langle z \rangle \approx \SI{10}{\nano\meter}$);
however, for all higher $E$, $1/T_{1}$ increases monotonically,
qualitatively consistent with the expected \gls{slr} response
from Meissner screening~\cite{2009-Hossain-PRB-79-144518,2012-Morenzoni-PRB-85-220501}. 
This rich behavior in the Meissner state is greatly contrasted by that in the normal state above $T_{c}$,
where any (meaningful) depth-dependence to $1/T_{1}$ is absent.
This fact is strong confirmation that the observed effects
originate from \ch{Nb}'s superconductivity.

With the essential features of the $1/T_{1}$ data described above,
we now consider a model that captures these details,
including the magnetic penetration depth $\lambda$.

\subsection{
	The $1/T_{1}$ Model
	\label{sec:results:model}
}

First, we remark that for a given \ch{^{8}Li^{+}} implantation energy $E$,
the measured $1/T_{1}$ represents an average over the probe's stopping profile
$\rho(z;E)$:
\begin{equation}
	\label{eq:depth-averaging}
	\frac{1}{T_{1}} \equiv \int_{0}^{\infty} \frac{1}{T_{1}(z)} \, \rho(z;E) \, \mathrm{d}z,
\end{equation}
where the term $1/T_{1}(z)$ encapsulates any depth dependence of the \gls{slr}
rate~\footnote{Note that the upper integration limit in \Cref{eq:depth-averaging} is justified by the fact that $\rho(z;E) \rightarrow 0$ for sufficiently large $z$.}.
While $\rho(z;E)$ can be reliably determined from Monte Carlo simulations
(see \Cref{sec:experiment,fig:srim}),
we must postulate a realistic form for $1/T_{1}(z)$.
In general,
there can be several contributions to the
\emph{observed} \gls{slr} rate $1/T_{1}$,
which can be decomposed into distinct mechanisms~\footnote{Assuming the fluctuations are statistically independent.}: 
\begin{equation*}
	\frac{1}{T_{1}} = \frac{1}{T_{1}^{e}} + \frac{1}{T_{1}^{d}} + \frac{1}{T_{1}^{q}} + \dots
\end{equation*}
Here, the $1/T_{1}^{i}$s denote contributions from:
conduction electrons ($i = e$),
dipole-dipole interactions ($i = d$),
and quadrupolar interactions ($i = q$).
While other contributions are possible~\cite{1961-Abragam-PNM,1983-Mehring-PHRNMRS,1990-Slichter-PMR,2019-Pell-PNMRS-111-1},
we restrict ourselves to the most common ones listed above.

We begin by considering $1/T_{1}^{e}$.
In metals, the \gls{slr} is often dominated by the so-called Korringa
mechanism~\cite{1950-Korringa-P-16-601,1990-Slichter-PMR},
wherein relaxation is induced by spin-flip scattering of conduction electrons
coupled to the nuclear spin via their hyperfine interaction.
From a derivation based on Fermi's golden rule, the sum over electron momenta,
when converted to an integral over energy, yields:
\begin{equation}
   \frac{1}{ T_{1}^{e} } = \frac{2\pi}{\hbar} A_{\mathrm{hf}}^{2} \int_{-\infty}^{+\infty} \rho_{e}^{2}(E) \, f(E) [ 1 - f(E) ] \, \mathrm{d}E,
\end{equation}
where $A_{\mathrm{hf}}$ is the hyperfine coupling,
$\rho_{e}(E)$ is the electronic \gls{dos},
and
$f(E)$ is the Fermi-Dirac distribution:
\begin{equation*}
	\label{eq:fermi-dirac}
	f(E) = \left [ \exp \left ( \frac{ E - E_{F} }{ k_{B} T }\right ) + 1 \right ]^{-1} ,
\end{equation*}
with $E_{F}$ and $k_{B}$ denoting the Fermi energy and Boltzmann constant,
respectively.
Note that the above integral is over all electron energies in the conduction band.
In a broad band degenerate metal,
where $E_{F} \gg k_{B} T$,
$\rho_{e}(E)$ is practically constant over the range where the Fermi factor
$f(E) [ 1 - f (E) ]$ is non-zero,
and the integral
(to an excellent approximation)
evaluates to:
\begin{equation}
   \label{eq:korringa}
   \frac{1}{T_{1}^{e}} \approx \frac{2\pi}{\hbar} A_{\mathrm{hf}}^{2} \, \rho_{e}^{2}(E_{F}) \, k_{B}T .
\end{equation}
The salient feature of \Cref{eq:korringa} is its linear temperature dependence,
which is distinct from most other \gls{slr} mechanisms.
This also produces an \gls{slr} rate that is \emph{independent} of
the applied magnetic field.
This electronic relaxation has been reported previously
for \ch{^{8}Li} implanted in this \ch{Nb} film~\cite{2009-Parolin-PRB-80-174109},
where below \SI{\sim 20}{\kelvin}:
\begin{equation*}
	\frac{1}{T_{1}^{e} T} \approx \SI{1.271 \pm 0.039 e-2}{\per\second\per\kelvin} .
\end{equation*}
Importantly, these rates are at least an order of magnitude smaller
than our observed $1/T_{1}$s,
leading us to conclude that the importance of $1/T_{1}^{e}$ is negligible
at our measurement conditions~\footnote{This is also consistent with the absence of a Hebel-Slichter coherence peak~\cite{1959-Hebel-PR-113-1504} in the $1/T_{1}$ data.}.

Next we consider $1/T_{1}^{q}$.
The \ch{^{8}Li} resonance in \ch{Nb} at low temperatures shows unambiguously that 
some probes stop in sites with non-cubic symmetry~\cite{2009-Parolin-PRB-80-174109},
where a static (i.e., time-averaged) \gls{efg} remains finite.
Considering \ch{Nb}'s \gls{bcc} structure~\cite{1965-Beshers-JAP-36-290},
there is some difficulty in precisely assigning a stopping site.
Earlier work suggested the possibility of an interstitial dumbbell-like \ch{^{8}Li-Nb}
defect complex at low-$T$~\cite{2009-Parolin-PRB-80-174109},
which is consistent with preliminary results from \gls{dft} calculations~\cite{Adelman-tbp};
however, no firm assignment has yet been made.
We note though that the presence of a non-zero \emph{static} \gls{efg}
does not necessitate a dynamic component sufficient for causing \gls{slr}
(cf.\ \ch{^{8}Li} \gls{bnmr} in \ch{Bi}~\cite{2014-MacFarlane-PRB-90-214422}).
Moreover,
$1/T_{1}^{q}$ is expected to follow a power-law that is strongly dependent
on temperature~\cite{1954-VanKranendonk-P-20-781},
but not applied field,
which is inconsistent with our data and measurements at higher fields~\cite{2009-Parolin-PRB-80-174109}.
At our measurement conditions,
such a phonon-mediated mechanism should be negligible,
given that $T$ is far below \ch{Nb}'s Debye temperature
(\SI{\sim 270}{\kelvin})
and there are very few thermally populated phonons.
On that basis,
we rule it out as a significant source for our
$1/T_{1}$~\footnote{In principle, this could be tested explicitly through a comparison of the \ch{^{8}Li} and \ch{^{9}Li} \gls{slr} rates (see e.g.,~\cite{2017-Chatzichristos-PRB-96-014307}).}.

With the above two mechanisms ruled out as meaningful contributors,
it is curious that a $T$-linear component of the \gls{slr} rate persists;
however, this is not a novel observation
and
similar behavior has been observed in the \ch{^{8}Li} \gls{bnmr} of materials
whose \gls{nmr} response is metallic
(see e.g.,~\cite{2018-MacFarlane-JPSCP-21-011020,2019-McFadden-PRB-99-125201,2020-McFadden-PRB-102-235206}).
As the exact origin remains unclear,
we adopted an empirical term in our model:
\begin{equation}
	\label{eq:linear}
	\frac{1}{T_{1}^{l}} \approx c \times T ,
\end{equation}
where $c$ is the slope of the linear relation.
This does not,
however,
account for the finite intercept at $T = \SI{0}{\kelvin}$,
which must find another origin such as the $1/T_{1}^{d}$,
considered below.

The \gls{slr} term $1/T_{1}^{d}$
accounts for relaxation resulting from magnetic fluctuations due to dipole-like
fields in the vicinity of the probe.
While the interaction is sensitive to the local field,
it is sufficiently decoupled by typical \gls{nmr} fields
(\SI{> 1}{\tesla})
such that its contribution to \gls{slr} is often minor.
Instead, the static (i.e., time-averaged) component of the dipole-dipole interaction
is often the main observable,
dictating the intrinsic width of the \gls{nmr} line~\cite{1948-VanVleck-PR-74-1168}.
Outside this ``high field'' limit, however,
both this quantity~\cite{1977-Hartmann-PRL-39-832}
and the \gls{slr} rate become field-dependent,
the latter often becoming dominant in \ch{^{8}Li} \gls{bnmr} 
(see e.g.,~\cite{2008-Parolin-PRB-77-214107,2009-Parolin-PRB-80-174109,2009-Hossain-PRB-79-144518,2012-Morenzoni-PRB-85-220501,2012-Chow-PRB-85-092103,2018-MacFarlane-JPSCP-21-011020}).
This contribution to the \gls{slr} can be considered as a form of
``cross-relaxation'' with the host \ch{^{93}Nb} nuclear spins
($I = 9/2$;
$\gamma/(2\pi) = \SI{10.4523 \pm 0.0005}{\mega\hertz\per\tesla}$;
\SI{100}{\percent} abundance)~\cite{2011-Baglin-NDS-112-1163},
whose dynamic behavior is stochastic with characteristic time-constants
$T_{1}$ and $T_{2}$ for their longitudinal and transverse components,
respectively
(the latter being of the greatest importance here).
In essence, the polarization of the \ch{^{8}Li} probes ``leaks''
into the \ch{^{93}Nb} spin ``bath'' at a rate $1/T_{1}^{d}$.
This process is strongly suppressed by large magnetic fields,
except near (avoided) level-crossings in the eigenstates of the coupled
spin system (see e.g.,~\cite{1980-Fujara-ZPB-37-151,2012-Chow-PRB-85-092103})
or at low magnetic fields where it dominates~\footnote{Alternatively, one may consider the low-field behavior as a form of avoided level-crossing resonance, with the ``universal'' resonance occurring at zero magnetic field.}.
In fact,
cross-relaxation between \ch{^{8}Li} and \ch{^{93}Nb} nuclei
has been studied previously using (neutron activated) \ch{^{8}Li} \gls{bnmr}
of \ch{LiNbO3}~\cite{1980-Fujara-ZPB-37-151}.

Qualitatively, the \gls{slr} behavior
of this cross-relaxation process
can be described by a Lorentzian-like expression,
with $1/T_{1}^{d} \propto B^{-2}$
(see e.g.,~\cite{2008-Parolin-PRB-77-214107,2009-Parolin-PRB-80-174109,2009-Hossain-PRB-79-144518,2012-Morenzoni-PRB-85-220501,2012-Chow-PRB-85-092103,2018-MacFarlane-JPSCP-21-011020}).
Recognizing that our \ch{^{8}Li} probe is unlike any of the ``host'' \ch{^{93}Nb} spins,
it is possible to describe their interaction more rigorously.
Explicitly,
for a probe spin $I$ coupled to a lattice spin $S$,
the heteronuclear dipole-dipole contribution to the \gls{slr} rate
can be written as~\cite{1983-Mehring-PHRNMRS}:
\begin{widetext}
\begin{equation}
	\frac{1}{ T_{1}^{d} } = \frac{1}{3} S ( S + 1 ) \langle \delta B_{d}^{2} \rangle \left \{ \frac{1}{3} J_{0}( \omega_{S} - \omega_{I} ) + J_{1}( \omega_{I} ) + 2 J_{2}( \omega_{S} + \omega_{I} ) \right \} ,
	\label{eq:dipole}
\end{equation}
\end{widetext}
where $\langle \delta B_{d}^{2} \rangle$ is the mean-squared fluctuating
field~\footnote{Note that the magnitude of $\langle \delta B_{d}^{2} \rangle$ is dependent on the \ch{^{8}Li} lattice site. Similarly, in environments where the inter-nuclear orientation in fixed (e.g., crystalline solids), it also depends on the applied field direction (see e.g.,~\cite{1961-Abragam-PNM,1983-Mehring-PHRNMRS,1990-Slichter-PMR}).}
and
\begin{equation}
	\label{eq:spectral-density}
	J_{n} ( \omega_{i} ) = \frac{ \tau_{c} }{1 + \omega_{i}^{2} \tau_{c}^{2} } ,
\end{equation}
which is the $n$-quantum spectral density function
(i.e., the Fourier transform of the (auto)correlation function describing
stochastic fluctuations in the local electromagnetic field that ``relax''
the probe's spin) 
characterized by an (exponential)~\footnote{This assumes that the \ch{^{93}Nb} spins have adopted a single spin temperature.} correlation time
$\tau_{c}$~\cite{1948-Bloembergen-PR-73-679}.
Note that while \Cref{eq:spectral-density} is among the simplest forms for $J_{n}$, most reasonable choices for the (auto)correlation function yield similar results (see e.g.,~\cite{1988-Beckmann-PR-171-85}).
In fact,
the validity of this formulation of heteronuclear dipole-dipole \gls{slr}
has been confirmed quantitatively in a previous \gls{bnmr} experiment on dilute
\ch{^{8}Li} in (isotopically pure) \ch{^{7}Li} metal~\cite{1985-Heitjans-JPFMP-15-41}.

In \Cref{eq:dipole,eq:spectral-density},
the connection to the local magnetic field is through the Larmor relation:
\begin{equation}
	\omega_{i} = \gamma_{i} B ( z ) ,
	\label{eq:larmor}
\end{equation}
where $\gamma_{i}$ is the gyromagnetic ratio of the $i$\textsuperscript{th} nuclear spin,
and $B(z)$ is the local field,
written suggestively to emphasize its spatial
(i.e., depth)
dependence.
With \Cref{eq:larmor} providing the connection
between the \gls{slr} rate
[via \Cref{eq:dipole,eq:spectral-density}]
and the local field,
we now consider the details of the latter.

In the normal state (i.e., when $T > T_{c}$),
there is no meaningful depth-dependence and $B(z) = B_{0}$,
the applied magnetic field.
In the limit that $\tau_{c}$ is temperature-independent,
\Cref{eq:dipole} reduces to a constant,
equivalent to the non-zero $1/T_{1}$ at \SI{0}{\kelvin}
(i.e., a finite intercept)
that we infer from extrapolating our data. 
By contrast, in the Meissner state the local field is expected to
be a strong function of $z$
and we ascribe the $E$-dependence of our data to this fact.
A salient feature of a superconductor in the Meissner state is
the complete expulsion of magnetic flux from its interior.
This is achieved by supercurrents within the material counter-acting
the external field,
causing it to ``decay'' over very short distances below the material's surface
(on the order of \SIrange{\sim 10}{\sim 100}{\nano\meter}).
Quite generally, the screening of the magnetic field can be written
as~\cite{1953-Pippard-PRSLA-216-547,1957-Bardeen-PR-108-1175,1996-Tinkham-IS-2}:
\begin{equation}
	B(z) = B_{0} \left ( \frac{2}{\pi} \right ) \int_{0}^{\infty} \frac{q}{q^{2} + K(q) } \sin( q z ) \, \mathrm{d}q ,
	\label{eq:pippard}
\end{equation}
where $B_{0}$ is the applied field, $z$ is the depth below the surface,
and $K(q)$ is the Fourier transform of the integrand kernel relating the
current density $\mathbf{j}$ to the vector potential $\mathbf{A}$ inside a
superconductor
(see e.g.,~\cite{2005-Suter-PRB-72-024506}).
In the absence of any appreciable dependence on the wavevector $q$,
$K(q) \approx 1 / \lambda^{2}$ and \Cref{eq:pippard} reduces to:
\begin{equation*}
	B(z) = B_{0} \exp \left ( - \frac{z}{\lambda} \right ) ,
	\label{eq:london}
\end{equation*}
which is the famous (phenomenological) result of the London
brothers~\cite{1935-London-PRSLA-149-71}.
A similar calculation,
accounting for the boundary conditions specific to a film with thickness $\zeta$,
as well as a non-superconducting ``dead layer'' $d$ at the material's surface
(see e.g.,~\cite{2023-McFadden-PRA-19-044018}),
yields~\cite{1996-Tinkham-IS-2}:
\begin{equation}
	B(z) = B_{0} \times \begin{cases} 
		1 , & z \leq d , \\
		\dfrac{ \cosh \left [ (\zeta / 2 - z - d) / \lambda \right ]  }{ \cosh \left [ \zeta / ( 2 \lambda ) \right ] } ,  &  d < z \leq \zeta .
	\end{cases}
	\label{eq:london-film}
\end{equation}
We note that \Cref{eq:london-film}
assumes no contribution from
non-local electrodynamics~\cite{1953-Pippard-PRSLA-216-547,1957-Bardeen-PR-108-1175,2005-Suter-PRB-72-024506}.
Though non-local effects are expected to be present in ``clean'' \ch{Nb}~\cite{2005-Suter-PRB-72-024506},
they are known to be weak and,
as is shown below, they can be safely neglected.

While \Cref{eq:dipole,eq:spectral-density,eq:larmor,eq:london-film} define
the connection between the \gls{slr} rate and
the spatial dependence of $B$,
these alone cannot account for the temperature dependence in $1/T_{1}$
that we observe.
To reconcile this deficiency,
it is necessary to also consider the temperature dependence of
the penetration depth $\lambda$.
Typically,
$\lambda$ is nearly temperature-independent when $T \ll T_{c}$,
but diverges such that $\lim_{T \to T_{c}} \lambda(T) = +\infty$. 
While this behavior arises naturally from \Cref{eq:pippard} in the \gls{bcs}
formulation of $K(q)$~\cite{1957-Bardeen-PR-108-1175},
the expressions are computationally cumbersome~\cite{2005-Suter-PRB-72-024506},
and we instead adopt the empirical ``two-fluid'' expression to describe the
temperature dependence:
\begin{equation}
	\label{eq:two-fluid}
	\lambda ( T ) =
	\begin{cases}
		\dfrac{ \lambda_{0} }{ \sqrt{ 1 - \left ( T / T_{c} \right )^{4} } } , & T < T_{c} , \\
		+\infty , & T \geq T_{c} ,
	\end{cases}
\end{equation}
where $\lambda_{0}$ is the penetration depth at \SI{0}{\kelvin}.
We note that \Cref{eq:two-fluid} has been shown previously
to accurately capture the temperature dependence of $\lambda(T)$
in \ch{Nb}~\cite{2005-Suter-PRB-72-024506}.

While these details can correctly explain the majority of our observations about the data,
a remaining feature requiring explanation is the upturn in $1/T_{1}$ observed at low-$T$
and low-$E$.
At these $E$ (\SI{< 2.5}{\kilo\electronvolt}),
we expect a substantial fraction of the implanted \ch{^{8}Li^{+}} stop within
the first \SI{\sim 5}{\nano\meter} of the surface
(see \Cref{fig:srim}),
where a native oxide layer
(e.g., \ch{Nb2O5})
resides~\cite{1987-Halbritter-APA-43-1}.
In the bulk,
these oxides are known to be paramagnetic
(e.g., due to oxygen sub-stoichiometry)~\cite{1991-Cava-PRB-44-6973},
and this magnetism has been observed at the
surface of \ch{Nb}~\cite{2005-Casalbuoni-NIMA-538-45,2011-Proslier-IEEETAS-21-2619,2021-Samsonova-IEEETAS-31-7000205}.
This leads us to postulate that the upturn is due to fluctuations in these
electronic moments localized in the surface oxide layer.
Noting that the upturn is appreciable at \SI{3.5}{\kelvin},
but negligible at \SI{11.9}{\kelvin}
(see \Cref{fig:slr-rates-film}), 
this \gls{slr} mechanism must be highly temperature dependent,
with the fluctuation spectrum ``freezing'' into
the vicinity of \ch{^{8}Li}'s Larmor frequency within a narrow temperature range.
While this process can be described, in general,
by a formalism analogous to that in \Cref{eq:dipole,eq:spectral-density},
our data is too limited for such a treatment and we instead
adopt an additional power law term for stopping depths less than \SI{\sim 5}{\nano\meter}:
\begin{equation}
	\frac{1}{T_{1}^{p}} \approx \frac{C}{ T^{b} } ,
	\label{eq:Nb2O5}
\end{equation}
where $b$ is the exponent and $C$ is a constant.
As is shown below,
this gives a reasonable description of our data.

Based on the above discussion,
we propose the following hierarchy to the \gls{slr} rates:
\begin{equation*}
	\frac{1}{T_{1}^{q}} < \frac{1}{T_{1}^{e}} \ll \frac{1}{T_{1}^{l}} \sim \frac{1}{T_{1}^{p}} \sim \frac{1}{T_{1}^{d}} .
\end{equation*}
We stress that only $1/T_{1}^{d}$ can account for the \gls{slr}
response observed upon transition from the normal to the
Meissner state~\cite{2009-Hossain-PRB-79-144518,2012-Morenzoni-PRB-85-220501}.
With these pieces in place,
we now turn our attention to the \ch{^{8}Li^{+}} stopping profile $\rho(z;E)$
---
the remaining ``ingredient'' required to evaluate \Cref{eq:depth-averaging}.

As mentioned in \Cref{sec:experiment},
\ch{^{8}Li^{+}} stopping profiles can be simulated reliably using Monte Carlo codes
(e.g., \gls{srim}~\cite{srim,2010-Ziegler-NIMB-268-1818}),
producing histograms to represent $\rho(z;E)$
(see \Cref{fig:srim}).
For our analysis,
it was convenient to have the ability to describe these profiles at
\emph{arbitrary} $E$, which can be accomplished by fitting the simulated
profiles and interpolating their ``shape'' parameters.
Empirically,
we found that $\rho(z;E)$ at a given $E$ can be described, in general, by:
\begin{equation}
	\label{eq:stopping}
	\rho (z; E) = \sum_{i}^{n} f_{i} p_{i} (z) ,
\end{equation}
where
$p_{i}(z)$ is a probability density function, 
$f_{i} \in [0, 1]$ is the $i^{\mathrm{th}}$ stopping fraction,
constrained such that
\begin{equation*}
	\sum_{i}^{n} f_{i} \equiv 1 ,
\end{equation*}
and $z$ is the depth below the surface.
For our \ch{Nb2O5}(\SI{5}{\nano\meter})/\ch{Nb} target (see e.g.,~\cite{1987-Halbritter-APA-43-1}),
the stopping data are well-described using $n = 2$ and a $p(z)$ given by
a modified beta distribution~\cite{2004-Gupta-HBDA}.
Explicitly,
\begin{equation}
   \label{eq:beta-pdf}
   p (z) =
   \begin{cases}
         0, & z < 0 \\
         \dfrac{ \left ( z / z_{0} \right )^{\alpha -1}  \left (1 -  z / z_{0} \right )^{\beta - 1}  }{ z_{0} \, B ( \alpha, \beta ) } , & 0 \leq z \leq z_{0} \\
         0, & z > z_{0}
      \end{cases}
\end{equation}
where $z \in [0, z_{0}]$ is the depth below the surface
and $B ( \alpha, \beta )$ is the beta function:
\begin{equation*}
   B ( \alpha, \beta ) \equiv \frac{ \Gamma (\alpha ) \Gamma (\beta) }{ \Gamma ( \alpha + \beta ) } ,
\end{equation*}
with $\Gamma (s)$ denoting the gamma function:
\begin{equation*}
	\Gamma (s) \equiv \int_{0}^{\infty} x^{s-1} \exp (-x) \, \mathrm{d}x .
\end{equation*}
Note that the ``extra'' $z_{0}$ in the denominator of \Cref{eq:beta-pdf}
ensures proper normalization of $p(z)$.
This (phenomenological) parameterization of the
\ch{^{8}Li^{+}} stopping profiles was found to work well for
$E \leq \SI{30}{\kilo\electronvolt}$
and a subset of the fit results are shown in \Cref{fig:srim}.

With all the pieces now in hand,
we briefly summarize the most important details of our $1/T_{1}$ model.
The major \gls{slr}
term $1/T_{1d}$ [\Cref{eq:dipole,eq:spectral-density}] has a field dependence of the form:
\begin{equation*}
	1/T_{1d} \propto 1/ \left ( 1 + \varsigma B^{2} \right ) ,
\end{equation*}
where $\varsigma$ is a scalar value.
Above $T_{c}$,
$B$ is equal to the external field and is independent of \ch{^{8}Li}'s position $z$
below \ch{Nb}'s surface,
resulting in a constant $1/T_{1d}$.
Below $T_{c}$, the local magnetic field in \ch{Nb} decreases as $z$ approaches the film's center
[\Cref{eq:london-film}]
due to the Meissner effect,
causing the spatially averaged $1/T_{1d}$ to increase with decreasing temperature.
This change is abrupt,
occurring immediately below $T_{c}$,
where the penetration depth $\lambda$ changes rapidly [\Cref{eq:two-fluid}].
At lower temperatures,
$\lambda$ approaches a constant value and so too does $1/T_{1d}$.
Similarly,
as the implantation energy $E$ is increased,
the implanted \ch{^{8}Li} sample depths closer to the film's center
(see \Cref{fig:srim})
where the local field is lowest,
causing the spatially averaged $1/T_{1d}$ to increase with increasing implantation energy
(up to the maximum $E$ of \SI{20}{\kilo\electronvolt} used here).

Finally,
with these details in mind,
it is straightforward to construct a model to describe the
$1/T_{1}$ data for \ch{Nb} in \Cref{fig:slr-rates-film}.
Explicitly,
a full \gls{3d} fit
(i.e., $1/T_{1}$ vs.\ $E$, $T$, and $B$)
was performed by:
combining
\Cref{eq:linear,eq:dipole,eq:spectral-density,eq:larmor,eq:london-film,eq:two-fluid,eq:Nb2O5}
into an expression for $1/T_{1}(z)$;
parameterizing $\rho(z;E)$
using
\Cref{eq:stopping,eq:beta-pdf} and interpolating their ``shape'' parameters;
plugging these expressions into \Cref{eq:depth-averaging}
and evaluating the integral numerically~\footnote{Note that, for the numeric evaluation of \Cref{eq:depth-averaging}, both adaptive Gaussian~\cite{quadpack} and tanh-sinh~\cite{1974-Takahashi-PRIMSKU-9-721,2001-Mori-JCAM-127-287} quadrature were found to work well.}.
The fit included \emph{all} experimental data below \SI{15}{\kelvin}~\footnote{This restriction ensures that the approximation in \Cref{eq:linear} holds; at higher temperatures the linear relation will likely break down due to \ch{^{8}Li} undergoing a site-change-transition (see~\cite{2009-Parolin-PRB-80-174109}).},
and the result is shown in \Cref{fig:slr-rates-film}.
A summary of the main fit parameters determined from this
procedure is given in \Cref{tab:results}.

\section{
	Discussion
	\label{sec:discussion}
}

It is clear from \Cref{fig:slr-rates-film}
that our model for the \ch{^{8}Li} $1/T_{1}$
(described in \Cref{sec:results:model})
is able to accurately reproduce our \gls{slr} data in the \ch{Nb} film.
Such good agreement is a strong confirmation of the choices made
throughout our analysis.
To further contextualize the results,
we now consider the extracted parameters explicitly.

We shall start our discussion with the film's critical temperature
$T_{c} = \SI{8.775 \pm 0.014}{\kelvin}$.
which is slightly below its ``bulk'' value of \SI{\sim 8.9}{\kelvin}
determined from magnetometry.
This difference, however, is consistent with the well-known suppression
of $T_{c}$ by an applied field
and its small uncertainty emphasizes that it is well-defined
by our model.
This has bearing on $\lambda(T)$,
whose temperature dependence is defined by \Cref{eq:two-fluid}.
This expression's other term of consequence,
the magnetic penetration depth at \SI{0}{\kelvin} $\lambda_{0}$,
was determined to be
\SI{51.5 \pm 2.2}{\nano\meter}.
The magnitude of this value is common for \ch{Nb} films
(see e.g.,~\cite{1981-Epperlein-PBC-108-931,1995-Andreone-PRB-52-4473,1995-Zhang-PRB-52-10395,1998-Pronin-PRB-57-14416,2005-Gubin-PRB-72-064503,2009-Nazaretski-APL-95-262502,2018-Pinto-SR-8-4710})
and larger than the so-called London penetration depth $\lambda_{L}$
often quoted for bulk \ch{Nb}
(see below).
Noting that our film is thick compared to \ch{Nb}'s \gls{bcs}
coherence length $\xi_{0} \approx \SI{40}{\nano\meter}$
(derived from an average of literature values~\cite{1965-Maxfield-PR-139-A1515,1966-Finnemore-PR-149-231,1968-French-C-8-301,1973-Auer-PRB-7-136,1974-Varmazis-PRB-10-1885,1981-Donnelly-PVM-118,1981-Epperlein-PBC-108-931,1991-Weber-PRB-44-7585,1992-Wood-NIMA-314-86,1995-Andreone-PRB-52-4473,1998-Pronin-PRB-57-14416}),
this ``discrepancy'' can be understood not in terms of a finite-size effect~\cite{2018-Pinto-SR-8-4710},
but rather in terms of the ``dirty''~\cite{1996-Tinkham-IS-2,2013-Dressel-ACMP-2013-104379}
character of our film.
It can be shown that, at \SI{0}{\kelvin}, one obtains the simple relationship~\cite{1996-Tinkham-IS-2}:
\begin{equation}
	\lambda_{0} = \lambda_{\mathrm{L}} \sqrt{1 + \frac{\xi_{0}}{\ell} } ,
	\label{eq:penetration}
\end{equation}
defining the influence of the carrier mean-free-path $\ell$ on $\lambda_{0}$.
Taking $\lambda_{L} \approx \SI{29}{\nano\meter}$
(based on an average of literature values~\cite{1965-Maxfield-PR-139-A1515,1966-Finnemore-PR-149-231,1968-French-C-8-301,1973-Auer-PRB-7-136,1974-Varmazis-PRB-10-1885,1981-Epperlein-PBC-108-931,1984-Felcher-PRL-52-1539,1991-Weber-PRB-44-7585,1992-Korneev-PSPIE-1738-254,1994-Kim-JAP-75-8163,1995-Andreone-PRB-52-4473,1995-Zhang-PRB-52-10395,1998-Pronin-PRB-57-14416}),
and re-arranging \Cref{eq:penetration},
we obtain
$\ell = \SI{18.7 \pm 2.9}{\nano\meter}$,
in good agreement with the range of values found in similarly prepared films
(see e.g.,~\cite{1972-Mayadas-JAP-43-1287,2001-CanizoCabrera-MPLB-15-639,2017-Broussard-JLTP-189-108,2018-Pinto-SR-8-4710}).
Within the Pippard~\cite{1953-Pippard-PRSLA-216-547} or \gls{bcs}~\cite{,1957-Bardeen-PR-108-1175} theories,
such a short $\ell$ corresponds to an ``effective'' coherence
length $\xi \approx \SI{13}{\nano\meter}$,
placing the results firmly within the ``local'' regime
(i.e., where $\lambda > \xi > \ell$ --- see e.g.,~\cite{2013-Dressel-ACMP-2013-104379}).
Note that, in this limit,
accounting for non-local electrodynamics in the calculation of $B(z)$~\cite{1953-Pippard-PRSLA-216-547,1957-Bardeen-PR-108-1175}
is unnecessary,
in contrast to that expected for ``clean'' \ch{Nb}~\cite{2005-Suter-PRB-72-024506}.

We now turn our attention to the correlation time $\tau_{c} = \SI{6.1 \pm 0.5 e-6}{\second}$
of the fluctuations causing the \ch{^{8}Li} \gls{slr}.
We remark that this value is larger by a factor of \num{\sim 10} than the quantity derived from the
\SI{\sim 0.2}{\tesla} ``width'' of the Lorentzian field-dependence  
quoted for earlier measurements at \SI{10}{\kelvin} and higher fields~\cite{2009-Parolin-PRB-80-174109}.
We emphasize that,
though not explicitly discussed in Ref.~\citenum{2009-Parolin-PRB-80-174109},
the field-dependence is a manifestation of the cross-relaxation discussed
in \Cref{sec:results:model}.
It should be remarked, however, that
differences in the two quantities are expected, for example,
due to the orientation dependence of spectral densities
(see e.g.,~\cite{1961-Abragam-PNM,1983-Mehring-PHRNMRS,1990-Slichter-PMR}).
While the earlier estimate was made using a limited set of measurements~\cite{2009-Parolin-PRB-80-174109},
our value is derived using both
a more comprehensive dataset
and
a more accurate model for the field dependence~\cite{1983-Mehring-PHRNMRS,1985-Heitjans-JPFMP-15-41},
suggesting that is the more reliable quantity.
This is confirmed by its close proximity to the
\ch{^{93}Nb} \gls{ssr} time $T_{2} \approx \SI{24}{\micro\second}$~\cite{1965-Butterworth-PPS-85-735},
differing from $\tau_{c}$ by only a factor of \num{\sim 4}.
Such a level of agreement is remarkable given the (indirect) manner
in which $\tau_{c}$ is identified.
We also note that
similar behavior has been observed for the low-field
\ch{^{8}Li} \gls{slr} in \ch{Bi2Te2Se}~\cite{2019-McFadden-PRB-99-125201}.
Interestingly,
the $\tau_{c}$ in the \ch{Nb} film is substantially shorter than in \ch{NbSe2}~\cite{2009-Hossain-PRB-79-144518},
presumably due to the reduced \ch{^{93}Nb} spin density and \ch{^{8}Li}'s weaker coupling to
their fluctuations.

Next,
we consider the ``coupling'' term in \Cref{eq:dipole}:
the dipolar field $B_{d} = \SI{41.1 \pm 2.1}{\micro\tesla}$.
Interestingly,
this value is a factor of \num{\sim 10} smaller than
estimates obtained from lattice sums~\cite{1948-VanVleck-PR-74-1168}
for several plausible interstitial sites in \ch{Nb}'s \gls{bcc} lattice~\cite{1965-Beshers-JAP-36-290},
as well as the \ch{^{8}Li} resonance linewidths, both
reported previously~\cite{2009-Parolin-PRB-80-174109}.
This level of difference isn't unexpected,
as the $B_{d}$ in \Cref{eq:dipole} is a \emph{dynamic}
(i.e., fluctuating) quantity,
which need not be similar in magnitude to its \emph{static}
(i.e., time-averaged) counterpart.
In fact,
this level difference is comparable to the case of \ch{^{8}Li} in \ch{Cu},
where estimates of $B_{d}$ from detailed measurements of $1/T_{1}$
in the vicinity of \glspl{alc}~\cite{2012-Chow-PRB-85-092103}
are an order of magnitude smaller that those
inferred from lattice sums~\cite{1999-Ohsumi-HI-120-419}
and resonance linewidths~\cite{1999-Ohsumi-HI-120-419,2007-Salman-PRB-75-073405}.
We stress that $1/T_{1}$ \gls{alc} measurements
should provide the most accurate determination of $B_{d}$;
however,
much like the use of lattice sums~\cite{1948-VanVleck-PR-74-1168,1977-Hartmann-PRL-39-832}
they require a structural model of
the \ch{^{8}Li^{+}} site in the host material.
While we do not have an independent experimental verification of our $B_{d}$ at this time,
we suggest that its value is reasonable.
Key to verifying this hypothesis will be developing a detailed understanding of
the \ch{^{8}Li^{+}} defect site using quantum chemical models.
Current information from high-field \gls{bnmr} experiments~\cite{2009-Parolin-PRB-80-174109}
and
preliminary \gls{dft} calculations~\cite{Adelman-tbp}
suggest that
a \ch{Li-Nb} dumbbell-like defect complex is formed at the
low temperatures employed in these experiments.

The next quantity to discuss is the slope of the
linear term $c = \SI{0.697 \pm 0.011}{\per\second\per\kelvin}$
in \Cref{eq:linear}.
There is some difficulty interpreting this value.
First,
we reiterate that this is \emph{not} due to a
Korringa mechanism~\cite{1950-Korringa-P-16-601,1990-Slichter-PMR},
whose slope is orders of magnitude smaller~\cite{2009-Parolin-PRB-80-174109}
(see also \Cref{sec:results:model}).
The stark difference in these quantities implies that $c$ is highly dependent on $B_{0}$,
though the precise mechanism is unclear
(e.g., fluctuations in the \emph{longitudinal} component of the \ch{Nb} spin-system).
At the level of empiricism,
this behavior has also been observed in the \ch{^{8}Li} \gls{bnmr} of other metallic systems,
including:
\ch{Au}~\cite{2008-Parolin-PRB-77-214107, 2018-MacFarlane-JPSCP-21-011020};
\ch{Cu}~\cite{2012-Chow-PRB-85-092103};
\ch{NbSe2}~\cite{2009-Hossain-PRB-79-144518};
and
\ch{Bi2Ch3} (\ch{Ch} = \ch{Se}, \ch{Te})~\cite{2019-McFadden-PRB-99-125201,2020-McFadden-PRB-102-235206}.
For the latter materials,
it was suggested that the change in $c$ may be related to charge-carrier-freezing
in an applied field
(see e.g.,~\cite{1956-Yafet-JPCS-1-137,1969-Dyakonov-PR-180-813,2005-Storchak-PRB-71-113202});
however, its applicability to our elemental metal seems minimal.
Alternatively,
the behavior in \ch{NbSe2} was thought to be due to
fluctuations in the \ch{^{93}Nb} $T_{1}$~\cite{2009-Hossain-PRB-79-144518},
based on reasonable agreement between the estimated $\tau_{c}$
and the $1/e$ decay time for the \gls{slr} in the \gls{nqr} regime~\cite{1996-Ishida-JPSJ-65-2341}.
While this remains the most plausible explanation for the behavior in \ch{Nb},
making an equivalent comparison is hampered by the unavailability of \ch{^{93}Nb} \gls{slr}
measurements at comparable fields
(see e.g.,~\cite{1965-Butterworth-PPS-85-735,1965-Noer-PPS-86-309,1975-Fradin-PLA-51-269}).
To be more conclusive,
a systematic study of $c(B)$ seems necessary.
Such an investigation may be possible in the near future thanks to a newly
developed \gls{bnmr} instrument designed to operate from \SIrange{0}{200}{\milli\tesla}~\cite{2023-Thoeng-RSI-94-023305}.

Finally, we consider the parameterization of the near-surface behavior of $1/T_{1}$,
namely, the upturn in \gls{slr} in the Meissner state at $E < \SI{2.5}{\kilo\electronvolt}$
and
the extent of the (non-superconducting) ``dead layer''.
We note that there is some ambiguity in quantifying these features,
owing to the rather limited data at these implantation energies.
Empirically,
the low-$E$ increase in $1/T_{1}$ can be described by
the power law in \Cref{eq:Nb2O5} with a fixed exponent
$b = 4.5$ and 
$C = \SI{2.03 \pm 0.19 e3}{\per\second\kelvin\tothe{4.5}}$.
Note that it was not possible to identify unique solutions for $C$, $b$, and $d$
without fixing one of the parameters, due to their high degree of correlation.
We are not aware of any physical significance for the value of our exponent $b$ used in \Cref{eq:Nb2O5},
making the connection of $C$ to the magnetic properties
of \ch{Nb2O_{5-x}}~\cite{1991-Cava-PRB-44-6973} unclear.
While paramagnetism in \ch{Nb}'s surface oxide layer~\cite{2005-Casalbuoni-NIMA-538-45,2011-Proslier-IEEETAS-21-2619,2021-Samsonova-IEEETAS-31-7000205}
remains the likely source,
further measurements
(e.g., in a \ch{Nb2O5} film)
are necessary to be conclusive.
Despite these ambiguities,
our model determined the superconducting ``dead layer'' $d$
to be \SI{5.0 \pm 0.9}{\nano\meter} for our film,
in excellent agreement with the typical thickness of the oxide layer
that forms natively at the surface of \ch{Nb}~\cite{1987-Halbritter-APA-43-1}.
While this value also compares well with \gls{le-musr} measurements
on a similar film~\cite{2005-Suter-PRB-72-024506},
it is much smaller than typical values found 
in \ch{Nb} prepared for \gls{srf} applications~\cite{2014-Romanenko-APL-104-072601,2017-Junginger-SST-30-125013,2023-McFadden-PRA-19-044018}.
This difference is likely due to greater surface roughness in the \gls{srf} samples~\footnote{\gls{srf} \ch{Nb} is typically formed from polished material made of bulk ingots. Similarly, films used in \gls{srf} applications are not deposited on ultra-flat substrates.},
which is known to influence the extent of the ``dead layer''
(see e.g.,~\cite{2012-Lindstrom-PP-30-249,2014-Lindstrom-JEM-85-149,2016-Lindstrom-JSNM-29-1499}).
We emphasize though that this quantity is inherent to a given \emph{sample},
rather than being intrinsic to the material.
Further \gls{slr} measurements at low implantation energies,
including their temperature dependence,
are required for deeper insight.

\section{
	Conclusions
	\label{sec:conclusions}
}

We used \ch{^{8}Li} \gls{bnmr} to measure the Meissner screening profile in
\ch{Nb}(\SI{300}{\nano\meter})/\ch{Al2O3} thin film.
Expanding on an earlier characterization performed at high magnetic field~\cite{2009-Parolin-PRB-80-174109},
we measured the \gls{slr} rate $1/T_{1}$'s dependence on temperature and \ch{^{8}Li^{+}} implantation energy
(i.e., depth below the surface),
in an applied field of \SI{20}{\milli\tesla}
(parallel to the film's surface).
While measurements at constant temperature in the normal state showed
no depth-dependence,
the \gls{slr} rate was found to vary strongly with implantation energy,
increasing deeper below the surface.
These observations were complimented by temperature dependent measurements
at constant implantation energy,
showing a linear variation in the normal state
($1/T_{1} = \SI{0.697 \pm 0.011}{\per\second\per\kelvin} \times T$),
but with a pronounced ``kink'' at $T_{c} = \SI{8.775 \pm 0.014}{\kelvin}$
and a non-linear increase at lower temperatures where $1/T_{1} \propto B(z)^{-2}$,
consistent with the \ch{^{8}Li} \gls{slr} in other
superconductors~\cite{2009-Hossain-PRB-79-144518,2012-Morenzoni-PRB-85-220501}. 
From a fit of the data to a model accounting for these details
and the \ch{^{8}Li^{+}} stopping profile,
it was found that the Meissner screening is well-described by a simple London model
with a magnetic penetration depth
$\lambda_{0} = \SI{51.5 \pm 2.2}{\nano\meter}$
(extrapolated to \SI{0}{\kelvin}).
The large $\lambda_{0}$ compared to \ch{Nb}'s intrinsic
London penetration depth $\lambda_{L} \approx \SI{29}{\nano\meter}$
is consistent with a relatively short carrier mean-free-path
$\ell = \SI{18.7 \pm 2.9}{\nano\meter}$
often found in similarly prepared films~\cite{1972-Mayadas-JAP-43-1287,2001-CanizoCabrera-MPLB-15-639,2017-Broussard-JLTP-189-108,2018-Pinto-SR-8-4710}.

This work constitutes an important advance in the use of
\ch{^{8}Li} \gls{bnmr} to study materials rich in (stable)
nuclear spins at low magnetic fields.
Similarly, it provides a basis for understanding and modelling
the \gls{slr} response in other superconductors,
where measurements in this field regime are likely required
(i.e., to remain in the Meissner state).
For the specific case of studying \ch{Nb},
this work provides a foundation for analyzing and interpreting
results in samples with \emph{engineered} surfaces
(e.g., from mild baking~\cite{2004-Ciovati-JAP-96-1591,arXiv:1806.09824}
or
doping~\cite{2013-Grassellino-SST-26-102001,2017-Grassellino-SST-30-094004}),
where $\lambda$ may be spatially \emph{inhomogeneous} 
(see e.g.,~\cite{2014-Barash-JPCM-26-045702,2019-Ngampruetikorn-PRR-1-012015,2021-Lechner-APL-119-082601}).
There is even further interest in extending this study to fields on the order of
\ch{Nb}'s superheating field $B_{\mathrm{sh}} \sim \SI{200}{\milli\tesla}$~\cite{2017-Junginger-SST-30-125012,2018-Junginger-PRAB-21-032002},
where \gls{srf} cavities typically operate.
A new beamline and instrument have been developed for
this purpose~\cite{2023-Thoeng-RSI-94-023305}
and the measurement program is already underway.

\begin{acknowledgments}
	We thank:
	R.~Abasalti, D.~J.~Arseneau, S.~Daviel, B.~Hitti, and D.~Vyas for
	their excellent technical support;
	as well as D.~Fujimoto for useful discussions.
	This work was supported by \gls{nserc} Awards to:
	T.~J., R.~F.~K., and W.~A.~M.
\end{acknowledgments}

\section*{Author Declarations}

\subsection*{Conflicts of Interests}

The authors have no conflicts to disclose.

\subsection*{Author Contributions}

\textbf{Ryan~M.~L.~McFadden}: Formal Analysis (lead); Investigation (equal); Software (lead); Visualization (lead); Writing --- Original Draft Preparation (lead); Writing --- Review \& Editing (lead).
\textbf{Md~Asaduzzaman}: Formal Analysis (supporting).
\textbf{Terry~J.~Buck}: Investigation (equal).
\textbf{David~L.~Cortie}: Investigation (equal).
\textbf{Martin~H.~Dehn}: Investigation (equal).
\textbf{Sarah~R.~Dunsiger}: Investigation (equal).
\textbf{Robert~F.~Kiefl}: Conceptualization (equal); Funding Acquisition (equal); Investigation (equal). Writing --- Review \& Editing (supporting).
\textbf{Robert~E.~Laxdal}: Formal Analysis (supporting); Writing --- Review \& Editing (supporting).
\textbf{C.~D.~Philip~Levy}: Investigation (supporting); Resources (equal).
\textbf{W.~Andrew~MacFarlane}: Conceptualization (equal); Funding Acquisition (equal); Investigation (equal); Writing --- Review \& Editing (supporting).
\textbf{Gerald~D.~Morris}: Investigation (equal); Resources (equal).
\textbf{Matthew~R.~Pearson}: Investigation (supporting); Resources (equal).
\textbf{Edward~Thoeng}: Formal Analysis (supporting); Writing --- Review \& Editing (supporting).
\textbf{Tobias~Junginger}: Formal Analysis (supporting); Funding Acquisition (equal); Writing --- Review \& Editing (supporting).

\subsection*{Data Availability}

Raw data from the \gls{bnmr} experiments were generated at TRIUMF's \gls{cmms}
facility and are available for download from: \url{https://cmms.triumf.ca/}.
Derived data supporting the findings of this study are available from the
corresponding authors upon reasonable request.

\bibliography{references.bib,unpublished.bib}

\end{document}